\documentclass[preprint]{aastex}

\begin{document}

\title{Mid-Infrared Spectra of Be Stars}
\author{S. A. Rinehart\footnote{Now at Queen Mary \& Westfield College,
Mile End Road, London E1 4NS, United Kingdom, email at S.Rinehart@qmw.ac.uk}, J.~R. Houck, \& J.~D. Smith}
\affil{Cornell University}
\authoraddr{Center for Radiophysics and Space Research, Cornell University, Ithaca NY 14853}
\email{rinehart@tristan.tn.cornell.edu, jrh13@cornell.edu,
jdsmith@astrosun.tn.cornell.edu}

\begin{abstract}

We present the first medium-resolution ($R\sim 600$) mid-infrared
(8-13.3\micron ) spectra of 11 Be stars.  A large number of lines are
observed and identified in these spectra, including, as an example, 39
hydrogen recombination lines in the spectrum of $\gamma$~Cas.  In the
majority of our spectra, all of the observed lines are attributable to
hydrogen recombination.  Two of the sources, $\beta$~Lyr and MWC~349
also show emission from other species.  Both of these objects show
evidence of [Ne~II] emission, and $\beta$~Lyr also shows evidence of
He I emission.  We tabulate the effective line strength and line
widths for the observed lines, and briefly discuss the physical
implications of the observed line series.  We also use a simple model of
free-free emission to characterize the disks around these sources.

\end{abstract}


\singlespace

\section{Introduction}

Be stars are defined as B stars which display optical emission in one or more
Balmer lines of hydrogen, or which have displayed such emission at
some point in the past.  There are also several subdivisions of Be
stars, including the ``classical'' Be stars and Herbig Be stars.
Herbig Ae/Be stars are considered the intermediate mass (5-20
$M_{\sun}$) counterparts to T Tauri stars; they are pre-main sequence
stars surrounded by an optically thick, dusty disk remaining from
proto-stellar collapse.  From this disk arises forbidden line emission
and H$\alpha$ emission.  Classical Be stars, on the other hand, are
intermediate mass main-sequence objects, believed to have an extended
optically thin gaseous envelope.  While these two types of objects
share similar nomenclature, they have little in common outside of
their masses.

The sources presented in this paper are all classical Be stars, and
henceforth the term Be star refers to classical Be stars.  Classical
Be stars are also divided into two categories, ``non-shell'', or
``normal'' stars, and ``shell'' stars.  Shell stars display broad
emission wings in one or more of the hydrogen lines and narrow
absorption bands for ionized metals (Slettebak 1988), features not
observed in non-shell Be stars.  Several suggestions about the
relation between shell stars and normal Be stars have been made: this
relation is complicated by the fact that several stars, including
$\gamma$~Cas and Pleione (HR 1180), have been observed to change from
``normal'' spectra to ``shell'' spectra (Doazan \& Thomas 1982).
Hanuschik (1996) explains shell stars as Be stars which are observed
equator-on, with the transition between normal and shell spectra
explained by density variations in a concave equatorial disk viewed at
a small inclination angle ($\sim$15$^{\circ}$).  Alternatively, it has
been suggested that the shell phase is a common evolutionary phase of
Be stars (Slettebak, {\it et al.} 1992), involving the formation of a
gas shell outside the stellar envelope.  The formation of the gas
shell can explain the transition from normal to shell spectra, while
the dissipation of the gas shell causes the transition from shell to
normal spectra.

Optical spectroscopy has raised a number of questions about Be stars.
Individual spectral lines are found to be double peaked, with a ratio
of the violet peak flux to the red peak flux
(the V/R ratio) which differs from unity.  Further, this ratio and
the line strengths of the hydrogen Balmer lines have been found to be
time-variable in many Be stars (Telting {\it et al} 1993).

Near-infrared (NIR) spectroscopy has added to this puzzle.
Variability of line strengths and V/R ratios of the double-peaked
lines mirror observations in the optical (Hamann \& Simon 1989).  In
addition, a number of the hydrogen recombination lines which appear in
the NIR have anomalously high line strengths, and the NIR continuum
has been observed to vary with time (Dougherty \& Taylor 1994).

Ultraviolet observations complicate the problem further.  The strong
lines observed in the UV should arise from a fast, low-density wind
(Smith 1995).  This is at odds with the optical and NIR observations,
which require a relatively dense gas to produce the observed features.
A possible explanation for this discrepancy is discussed below.

X-ray observations have shown several Be stars to be X-ray
binaries.  $\gamma$~Cas, the brightest of the northern hemisphere Be
stars has a compact companion, most likely a white dwarf (Haberl
1995).  Other Be stars are also in binary systems; a large number of
Be stars, including many of the sources presented in this sample, are
spectroscopic binaries, including $\beta$~Lyr, $\zeta$~Tau, and
$\phi$~Per (Jared {\it et al.} 1989).

Of all of the wavelength regions, the mid-infrared has been the least
explored.  Photometry by Gehrz {\it et al} (1974) from 2.3 to
19.5\micron \ showed that the spectral energy distribution of the IR
excess was consistent with free-free emission from a warm disk-shaped
envelope.  It has also been suggested (when discussing NIR
observations) that studies of infrared line emission can be very
useful in probing the stellar envelope of Be stars, as the emission is
highly dependent upon the temperature and density structure of the
envelope (Marlborough {\it et al.} 1997).  This is one of the
motivations for the study introduced here.

Explaining the multitude of observations by modelling is a daunting
task.  The most common model used to explain the observations is the
axisymmetric equatorial disk model of Poeckert \& Marlborough (1978).
A slow, high-density wind at the equator creates a thick envelope
which produces the observed optical and NIR emission, while the UV
emission arises from a fast, tenuous wind originating at the poles of
the star (Lamers \& Waters 1987).  The presence of an equatorial disk
has been supported by interferometric and polarimetric observations
(Quirrenbach {\it et al.} 1997, Stee {\it et al.} 1995), but these
models are incomplete, as they still cannot explain some Be star
phenomenae, such as the observed V/R variability.

A more complete discussion of Be stars can be found in several
excellent review articles, including Slettebak (1988) and Dachs (1986).

In this paper, we present the first high-sensitivity, medium
resolution ($R\sim 600$) mid-infrared (8-13.3\micron ) spectra of be
stars.  In \S 2, we discuss the observations and data reduction
techniques.  In \S 3, we present the observed spectra, and briefly
mention characteristics of particular sources.  In \S 4, we discuss
using free-free emission from the stellar envelope to model the MIR
continuum.  Finally, in \S 5, we briefly discuss some of the
implications of these observations.

\section{Observations and Data Reduction}

The observations presented here were taken using SCORE (van Cleve
et al. 1998) at
the Cassegrain focus of the Hale 5-m telescope.\footnote{Observations at
the Palomar Observatory were made as part of a continuing collaborative
agreement between the California Institute of Technology, the Jet
Propulsion Laboratory, and Cornell University.}  SCORE is a novel medium
resolution mid-infrared spectrograph built originally as a proof of
concept of the Infrared Spectrograph (IRS) short-hi module for SIRTF.
It uses cross-dispersion to obtain a 7.5-15\micron \ spectrum in a single
exposure.  The obtained spectrum has a resolution of $\sim$600, with a
1$\arcsec \times 2\arcsec$ slit.  This short slit length is ideal for
sources which are unresolved in the MIR at Palomar.  SCORE also makes
use of a second MIR array for slit-viewing, which both simplifies source
acquisition during observing and provides information for photometric
calibration (see discussion below).  A listing of all of
the observations is presented in Table~1.  All of these observations
were made using standard beam-switched chopping and nodding techniques.
Typical integration times ranged from 240 to 1090 seconds, and the
entire N-band was observed at a single setting for each object.  The
1$\sigma$ sensitivity at 11.0\micron \ in 100 seconds of integration for
our instrument is roughly 90mJy.  The MIR seeing for these observations
was typically 1$\arcsec$.

The individual integrations of each object were stacked, and the
N-band spectra were extracted using SCOREX, extraction software
designed to optimally extract the spectrum from the cross-dispersed
format of SCORE (Smith et al. 1998).  These spectra were then calibrated
using observed spectra of standard stars (Cohen, {et al.} 1992) to
properly remove spectral features from the sky background.  The
observed spectra from these standard stars were taken at nearly the
same airmass as our observed objects, as even relatively small
differences in airmass result in different strengths of atmospheric
features, leading to inaccurate subtraction of background features
from our calibrated spectra.

Absolute photometry for the spectra was found by using the slit-viewer
images of SCORE.  Since a small chop amplitude ($\sim$5\arcsec ) was used
for these observations, the chopped images appear in the field-of-view
of the slit-viewer.  We calculate the slit-throughput by comparing the
flux incident on the slit-viewer focal plane with the total flux found
in the observed spectrum; mathematically,

\begin{equation}
A = \frac{F_{\nu}(\rm Slitviewer \mit )}{\int F_{\nu}(\lambda) g(\lambda)\ \rm d\mit\lambda}
\end{equation}

\noindent where $A$ is the renormalization factor, $F_{\nu}(\lambda)$
is the flux as a function of wavelength on the spectrograph,
$g(\lambda)$ is the slit viewer filter passband function, and
$F_{\nu}(\rm Slitviewer \mit )$ is the total flux incident on the
slit-view focal plane array.  The renormalization factor $A$ should
account for the slit throughput.  To provide the absolute calibration,
we use the same procedure for our calibrator star, deriving another
normalization factor, $A_{calib}$.  By multiplying the observed,
calibrated spectrum of our source by $\frac{A}{A_{calib}}$ we are able
to normalize the observed spectrum to have the correct amount of flux,
without introducing errors from instrumental effects.  The absolute
normalization of the spectra is uncertain to roughly 5\%, due
primarily to the uncertainty in the slit throughput efficiency.

\section{Observed Spectra of Be stars}

In Figure~1, we present the reduced spectra of our objects.  These
spectra clearly show the presence of a large number of spectral lines.
Further, in most of the spectra, there are a number of lines which are
weak relative to the strong MIR continuum.  Since these weak lines
appear in several different spectra, but not in calibration star
spectra, we conclude that they are not artifacts.  Further, the clear
presence of multiple lines in each series implies the presence of other
lines within the series.  For instance, the H 29$\rightarrow$10 is weak
in the spectrum of $\gamma$~Cas, but the presence of other transitions,
(including the H 29$\rightarrow$9 transition in this case) support the
identification of this weak line, since they arise from a common excited
state.  Similar support can be found for nearly all of the weak observed
lines.  Finally, because all of the lines occur at wavelengths
attributable to hydrogen recombination (for most of our objects, see
below), there is strong evidence to support the pure-hydrogen
recombination spectra.

There is also another test for the presence of these weak lines.  Nine
of the 11 spectra display similar strong hydrogen line emission (all
but $\beta$~Lyr and MWC~349), but a number of these spectra have low
signal to noise.  These nine sources will hereafter be referred to as
the hydrogen spectra Be stars (or HS stars).  If we suppose that all
of the lines observed in the high signal to noise spectra (such as
$\gamma$~Cas and $\zeta$~Tau) are present in the spectra of all HS
sources, we can coadd the spectra from all nine of these objects.  If
these lines are, in fact, present in a majority of the sources, we
expect the weak lines to be accentuated.  If, on the other hand, these
weak lines are present in only a few spectra, we expect them to be
diluted (relative to their appearance in the individual spectra).
This coadded spectrum is shown in Figure~2, with the line
identifications marked at the top of the plot.  Examination of this
spectrum shows a large number of lines, including a number which are
only observed as very faint peaks in the spectra of our bright
sources.  This supports the idea that all of the observed hydrogen
transitions are present in the majority of the HS sources. 

In Table~2, we list all of the observed lines, their line widths and
effective line strengths, and the hydrogen recombination transition
responsible for their production, in order of decreasing wavelength.
This table includes the data for all of the HS sources (which excludes
$\beta$~Lyr and MWC~349).  The data from the two peculiar objects
($\beta$~Lyr and MWC~349) is presented in Table~3, because of the
clear differences between their spectra and the spectra of the
majority of our sources.  Note that in Table~2, every observed line is
attributable to hydrogen recombination.  We examined the spectra for
the presence of line species other than hydrogen, in particular for
the presence of He I or He II, since these species can produce
emission lines coincident with hydrogen lines.  He I is easily
identified by the presence of the 10.88 \micron \ He I 3S--3P$_o$
transition, while He II can be identified by the presence of many
additional hydrogen-like transitions.  In the nine HS spectra, there
is no evidence for the He I 10.88\micron \ transition or for the
numerous He II transitions which would be present if the species
accounted for any of the observed emssion.

In $\beta$~Lyr, however, we observed the He I 3S--3P$_o$ transition at
10.88 \micron .  Since Helium recombinations occur at the same
wavelength as hydrogen recombinations (for instance, both H
9$\rightarrow$7 and He I 9$\rightarrow$7 occur at 11.3\micron )\, the
presence of this line implies that the observed lines are not
necessarily purely due to hydrogen, as is true for most of our other
sources.  The possible overlap of hydrogen and helium recombination line
makes exact identification of these lines problematic.  We also find
evidence for the [Ne~II] 12.8\micron \ transition, a line also observed
in the spectrum of MWC~349.  MWC~349 showed no evidence for He I or He
II emission, and only showed a few weak hydrogen transitions (with the
exception of the H 7$\rightarrow$6 transition which was quite strong).
We will return to a discussion of the observed spectral features of
these objects later in this section.

\subsection{Hydrogen Spectra Be Stars}

While the nine HS sources presented here have similar MIR spectra,
they are significantly different types of objects in several ways.
Seven of the sources are classified as shell stars.  Five of our
sources are spectroscopic binaries, and three of the sources are X-ray
binaries.  Further, while most Be stars do not display radio emission,
four of the sources presented here are relatively bright radio
sources.  The characteristics of the individual objects are tabulated
in Table~4.  For completeness, some of the characteristics of the two
peculiar sources are also listed on this table.

The large differences in the properties of the individual objects
leads us to look for correlations.  Are any of these individual
properties related to the MIR observations presented here?  We looked
at correlation plots for each of the different categories listed in
Table~4.  An example of one of these correlation plots is shown in
Figure~3.  We have plotted the ratio of the H 9$\rightarrow$7 to H
7$\rightarrow$6 transition versus the ratio of the H 10$\rightarrow$7
to H 7$\rightarrow$6 transitions in this figure, with the XRB plotted
as $+$ symbols, spectroscopic binaries as open triangles, and single
stars as open diamonds.  We have examined a number of similar
correlation plots, looking for evidence that our MIR spectra are
effected by each of these properties, but have found no such evidence.
However, given the small statistics (only nine sources), this lack of
detection is not surprising, and only by building a more significant
set of data will such correlation tests be valuable for understanding
the effects of these peculiarities on our MIR spectra.

The presence of so many high-level hydrogen transitions provides
valuable insight into the origin of line emission.  The line strengths
are inconsistent with optically thin line emission (Hummer \& Storey
1987), and therefore must originate at optical depths of $\sim$1.  The
optically thick emission will simply be the product of the Planck
function, the line width, and the surface area of the emission region.
Since the Planck function of a gas with temperatures consistent with the
production of such high H recombination series decreases with wavelength
and our observations show no clear trend in line width with wavelength,
the surface area must be {\em increasing} with wavelength.  ISO
observations have shown that the line width actually decreases with
wavelength (Hony et al. 1999), strengthening this statement.  This
implies a density gradient, in order to balance line ratios; if the
density is uniform, then the low-level transitions will become optically
thick much more quickly than their high-level counterparts, giving them
an effective emission surface significantly larger than the high-level
lines, and overcompensating for the decreasing strength with wavelength.
This was previously noted from NIR observations of $\gamma$~Cas (Hamann
\& Simon 1987).  These authors concluded that gradients as small as
r$^{-2}$ were sufficient to explain their observed line ratios.

\subsection{Peculiar Be Stars}

$\beta$~Lyr and MWC~349 display spectra which are significantly
different from our HS sources, and from each other.  Each of these
sources is unique, and it is therefore not surprising that their MIR
spectra should be peculiar as well.  We briefly discuss the peculiar
aspects of the MIR spectra of these objects here, and present some
possible explanations for these peculiarities.

$\beta$~Lyr is a known spectroscopic and eclipsing binary.  The two
members of this binary are B stars, with masses of $M_1 \approx 13
M_{\sun}$ and $M_2 \approx 3 M_{\sun}$ (Harmanec \& Scholz 1993).  The
two stars are in close orbit, with the mass-loss from the less-massive
star filling its Roche lobe, overflowing to form an accretion disk
around the more massive star (De Greve \& Linnell 1994).  This situation
is quite different from the disk-like stellar envelopes around our HS Be
stars, which occupy a comparatively small volume.  

The accretion process can also possibly explain observed He I emission
by invoking convective dredging in the donor star, thus bringing helium
into the Roche lobe and the accretion disk.  The presence of the He
recombination lines in this spectrum, when compared to the HS sources is
most easly explained by either a higher He abundance in the
circumstellar disk or a higher level of excitation.  If we compare
$\beta$~Lyr to $\gamma$~Cas, however, we see that $\gamma$~Cas is
significantly hotter than $\beta$~Lyr ($T_{\beta Lyr} < 20000$K).  Since
no He emission is observed in $\gamma$~Cas, whose disk gas is at higher
excitation, we conclude that the observed He emission in $\beta$~Lyr is
due to a greater He abundance.  

The biggest question raised from examination of the MIR spectrum of
$\beta$~Lyr is the cause of [Ne~II] emission.  While a great number of
emission lines have been observed in the optical and NIR spectra of
$\beta$~Lyr, there has been no evidence for forbidden emission in any of
these spectra (Johnson 1978), or for mid-infrared emission from
forbidden lines in our spectra ([Ar III] at 9.0\micron , [S IV] at
10.5\micron ).  Why should [Ne~II] emission, an no other forbidden
emission, be observed?  The two properties which should be examined in
particular are the ratio of the observed to critical density
($n_e/n_{crit}$) for [Ne~II] emission and the energy required to form
Ne~II.  Upon examination, we find that both of these characteristics of
Ne are favorable for the observed forbidden emission.  [Ne~II] has a
very high critical density ($10^5$ cm$^{-3}$), higher than any other
astronomically strong MIR forbidden transition.  Further, it has a
relatively low ionization potential (21.5 eV), making it possible to
ionize a large enough fraction of the Neon gas by the hot B stars to
create the observed emission.

Using the observed strength of the [Ne~II] line, the distance to
$\beta$~Lyr (270 pc) from the Hipparchos catalog (1997), and making a
few assumptions about the emitting gas (e.g. $n_e = n_{crit}$, $T_e =
10^4$K, solar abundances), we can calculate an approximate mass of
emitting gas.  This calculation gives a mass of the emitting gas of
$M_{tot} \approx 10^{-5}M_{\sun}$.  This compares favorably with
calculations of the mass-transfer rate from the ``donor'' star to the
``accretor'' star, if we assume that some small fraction of
mass-transfer escapes into an extended gas shell.  Such calculations,
assuming complete mass conservation between the two stars, give values
of $\dot{M} \approx 2\times 10^{-5}$ M$_{\sun}$ yr$^{-a}$ (Hubeny et
al. 1994).  If we assume that the gas is fully ionized, we can estimate
an emission volume for the [Ne~II] line.  We assume, for this
approximation, that the emission volume is spherical, and find that the
emitting region should be of order 100 AU across.  For comparison, the
separation of the two components of the binary is only of order 20 AU.
From these estimates, we suggest that the forbidden line emission does
not arise in the Roche lobe or accretion disk, but must come from a much
more extended gas shell around the binary.  This could also explain the
difference in the observed hydrogen spectra of this source, relative to
our HS sources.  The extended, low density outer gas shell will produce
optically thin line emission from hydrogen, while the disk produces
optically thick line emission.  The contribution from the optically thin
emission will alter the hydrogen line ratios, explaining why the line
ratios for hydrogen emission from $\beta$~Lyr are not $\sim$1 (as for
our HS sources).  

MWC~349 has the largest MIR to optical flux ratio of all of our sources,
with this ratio more than 100 times larger than our other sources.  This
could be due to several different phenomena, including a much cooler
source or much more circumstellar mass (including, perhaps, a large
amount of dust, leading to severe reddening).  Independent of the cause,
the red color of MWC~349, compared to our HS sources, implies that the
nature of this object or its environment may be quite different from
classical Be stars.  Observations of MWC~349 in the submillimeter
detected H$\alpha$ transitions which were greatly amplified; MWC~349 was
the first source observed with hydrogen laser emission (Martin-Pintado
{\it et al.}  1989).  Further, recent ISO observations have detected
more than 160 emission lines between 2.4 and 190\micron , including
every hydrogen $\alpha$ transition (H$5\rightarrow 4$ to H16$\rightarrow
15$) in this range (Thum {\it et al.} 1998), showing the presence of
infrared lasers from the large amplification of the H$\alpha$ lines.
Further, lasing/masing has been observed from a number of the hydrogen
$\beta$ lines.  It is thought that the conditions which allow the
observed masing/lasing from MWC~349 include the high temperature of (and
corresponding large ultraviolet flux from) the central star, a dense,
massive, neutral disk (much more massive than the disks in most Be
stars), and the coincidental edge-on view we have of the disk
(Strelnitski {\it et al.} 1996).

A significant amount of forbidden line emission in the optical has been
observed in MWC~349 (Brugel \& Wallerstein 1979, Andrillat \& Swings
1976, Allen \& Swings 1976)).  We have also detected forbidden emission,
in the form of the [Ne~II] 12.81\micron \ line.  It is unlikely that
this forbidden emission arises from the same region as the hydrogen
emission, however.  Recent work by Them \& Greve (1997) used the
observed Paschen decrement to estimate the electron density in the disk
at $n_e = 10^8$, over one hundred times greater than the critical
density for [Ne~II] forbidden emission.  The forbidden emission more
likely arises in an extended gas component around the star.  Such an
extended component has been observed via radio measurements and has been
associated with a slow (50 km s$^{-1}$) gaseous outflow.  Cohen et
al. (1985), using observations from the VLA, found that the radio
observations were well-matched by a spherical $1/r^2$ wind model, with
mass-loss rates of order 10$^{-5}$M$_{\sun}$ yr$^{-1}$ and a temperature of
roughly 9000 K.  

Using the same technique as described for $\beta$~Lyr, we estimate the
mass of the emitting gas from the forbidden emission.  We use the same
assumptions as in the case of $\beta$~Lyr, and estimate a distance to
the source of 400 pc (Thompson et al. 1977).  From these, we find a mass
of the emitting gas of $M_{tot} \approx 10^{-2}M_{\sun}$.  The mass of
the disk should be substantially larger than this, up to several
solar masses (Thompson et al., Thum \& Greve 1997).  However, the
forbidden emission we observe will only come from the regions around the
star where the density is relatively low, and therefore will only arise
from a small fraction of the disk, since the majority of the disk has a
high ($\sim 10^8$cm$^{-3}$) density (Thum et al. 1998).

\section{Continuum Emission}

The majority of photospheric radiation from hot stars is at short wavelengths, in the optical or ultraviolet (an
16000 K star has a peak flux at $\sim$ 1800\AA , well into the UV).  In
the MIR, photospheric emission is well inside the Raleigh-Jeans tail of
the blackbody distribution, producing relatively weak emission in this
wavelength range.  Classical Be stars have been long known to produce
large IR excesses, and a common explanation for this is emission via
free-free processes.

Free-free radiation from ionized gas is a powerful source of
emission at long wavelengths.  Typically, such emission has been used
to explain radio emission from cool astronomical sources (Panagia \&
Felli 1975), but it can also explain the infrared excesses observed
from many sources.  Be stars have well-known IR excess, and free-free
emission from the extended stellar envelope provides a convenient
explanation for this excess.

Gehrz {\it et al.} (1974) used models of optically thin and optically
thick free-free emission to match infrared photometry of Be stars from
2.3\micron \ to 19.5\micron .  Waters {\it et al} (1984) successfully
modeled IRAS Be star observations with an optically thick free-free
emission.  We have used models of free-free emission to fit our MIR
continuum observations, finding values for the shell radius ($R_{sh}$),
the electron density ($n_e$), and the shell temperature ($T_{sh}$).

\subsection{Calculations of Free-Free Emission}

Three emission processes are included in our model of the MIR
continuum: emission from the photosphere of the star, optically thin
free-free emission, and optically thick free-free emission.

Because neither of the free-free processes are very effective at short
wavelengths, we can constrain the stellar parameters without having to
consider the effects of the shell.  In the infrared, the photospheric
spectrum is essentially a blackbody produced at the photospheric
temperature, so the Planck function describes this emission.  Initial
values for the temperature ($T_{\star}$) and radii ($R_{\star}$) of our
sources were approximated from their respective spectral types (Waters
{\it et al.} 1987), and we then adjusted the values of both $T_{\star}$
and $R_{\star}$ to agree with J-band (1.25\micron) and K-band
(2.20\micron) photometry (Gezari {\it et al.} 1993).  Distances to the
stars were obtained from the Hipparchos Catalog (1997).  All of these
parameters, including the J-band and K-band fluxes, are listed in
Table~5.

To fit the observed MIR continuum, we require several assumptions.
First, we assume that the MIR emission arises from an extended stellar
envelope around the star.  Further, it is assumed that this envelope is
flattened into an oblate spheroidal disk, with a semi-minor axis
$\approx 1/10 R_{sh}$ where $R_{sh}$ is the semi-major axis of the
spheroid.  Further, we assume that the stellar envelope has both uniform
density and uniform temperature.  These assumptions are clearly not
correct, but greatly simplify calculations without introducing
significant amounts of error.  Finally, we assume that we are looking
into the disk-like envelope edge-on.  This last assumption is supported
by two facts: the large values of $v\sin i$ for these sources, and the
central line absorption of shell spectra (seven of our nine HS sources
are shell stars, see Table~4).

The MIR emission from the stellar photosphere is calculated using the
Planck function (making use of $R_{\star}$, $T_{\star}$, and $D$), as
described above, but is corrected by applying an extinction factor in
the form of $e^{-\tau (\rho=R_{sh})}$.  This will account for free-free
absorption of photospheric radiation as a function of wavelength.
Thus, we write

\begin{equation}
F_{\lambda}^{\star} = 37469.4\left[e^{\left(1.44 \frac{1 \mu\rm m\mit}{\lambda}\frac{10^4}{T_{\star}}\right)} - 1\right]^{-1} \left(\frac{1 \micron}{\lambda}\right)^5 \left(\frac{R_{\star}}{D}\right)^2 e^{-\tau(\rho=R_{sh})}
\end{equation}

Next, we include the effect of optically thick free-free emission.
The optically thick emission will be characterized by
the Planck function at the shell temperature ($T_{sh}$).  This emission will
arise from an area of the disk which is defined by the surface a depth
$\rho$ into the disk, where the optical depth is 

\begin{equation}
\tau^{f\hspace{-0.04cm}f} = \alpha_{\nu}^{f\hspace{-0.04cm}f}\rho = 1.4\times 10^{-3} \left(\frac{n_e}{10^{11}}\right)^2\left(\frac{10^4}{T_{sh}}\right)^{0.5}\left(\frac{\lambda}{\micron}\right)^3 \left( 1- e^{-1.44 \frac{\mu\rm m\mit}{\lambda}\frac{10^4}{T_{sh}}}\right) \left(\frac{\rho}{10^{12}}\right) = 1
\end{equation}

\noindent This cuts off the optically thick emission at short wavelengths, as
the absorption coefficient becomes very small and the entire disk
becomes optically thin.  Mathematically,

\begin{equation}
F_{\lambda}^{f\hspace{-0.04cm}f-thick} = 37469.4\left[e^{\left(1.44 \frac{1 \mu\rm m\mit}{\lambda}\frac{10^4}{T_{sh}}\right)} - 1\right]^{-1} \left(\frac{1 \mu\rm m \mit}{\lambda}\right)^5 \left(\frac{R_{sh}}{D}\right)^2 x(\tau)
\end{equation}

\noindent where $x(\tau)$ is the ratio of the area of the optically thick emission to the maximum area of optically thick emission.

Finally, we include the effect of the optically thin region of the
emission disk.  This can be accounted for to first order by
integrating the emission function for free-free emission
($\epsilon_{\nu}^{ff}$) over the volume of the stellar envelope.  We
improve upon this first-order calculation by also including the
extinction ($e^{-\tau(\rho)}$) factor in the integral over volume.  In
this way, we account for the reduction in optically thin emission with
increasing wavelength.  This gives us, replacing Equation~3,

\begin{equation}
F_{\lambda}^{f\hspace{-0.04cm}f-thin} = 7.12\times10^{-13}\left(\frac{n_e \ \mu\rm m\mit \ \rm pc\mit}{10^{11} \lambda D}\right)^2\left(\frac{10^4}{T_{sh}}\right)^{0.5}\left(\frac{R_{sh}}{10^{12}}\right)^3 \left(e^{-1.44\frac{1 \mu\rm m\mit}{\lambda}\frac{10^4}{T_{sh}}}\right)\left(\frac{\int e^{-\tau(\rho)} dV}{V_{full}}\right)
\end{equation}

To fit our observations, we sum the emission from these three sources,
varying the temperature ($T_{sh}$), radius ($R_{sh}$), and electron
density ($n_E$) of the shell.  This sum is then compared to our observed
spectrum.  We calculate a reduced $\chi^2$ value based upon nine points
in our observed spectra; these nine points located in regions at least
0.05\micron \ from any of the observed emission lines, in order to not
contaminate the $\chi^2$ with line emission.  The flux at each point is
determined by averaging the flux over $\lambda -0.05$ to $\lambda
+0.05$\micron , in order to reduce the effects of local noise spikes.
The results of these fits are shown in Table~6.  In Figure~4, we show
the observed spectrum of $\gamma$~Cas with the overlaid fit.  The
observed spectrum is shown as a dotted line, the fit curve is shown as a
solid line, the dashed line and dot-dash line are the optically thin and
optically thick components of the emission respectively, the diamonds
are the nine points used to calculate the reduced $\chi^2$, and the open
triangle is the 12\micron \ flux measured by IRAS.

\subsection{Comparisons with Previous Work}

Two works in particular, Gehrz {\it et al.} 1974 and Waters {\it et al.}
1987, have calculated parameters of Be stellar envelopes.  Gehrz {\it et
al.} calculated $R_{sh}$ and $n_e$ from their observations, based upon
an estimate of $T_{sh}$ (this estimate was constrained solely by the
requirement that $T_{sh}$ be between 10000K and $T_{\star}$) by using
the estimated $T_{sh}$ and by calculating the wavelength where the
extrapolations of optically thin and optically thick free-free emission
intersect (at which point the optical depth $\tau (\lambda _{c}) \equiv
1$).  This provides a relation between the radius of the disk and the
electron density.  Further, by assuming a Rayleigh-Jeans distribution
for both the optically thick shell flux and the stellar flux (at
$\lambda _{c}$), they obtain an expression for the radius of the shell,
which then determines the electron density of the disk.  For these
calculations, they have assumed the disk to be flattened with an aspect
ratio $A = r_{s}/d_{disk} = 5$ (where $d_{disk}$ is the full thickness
of the disk), the same value used for our modelling.  As in our model,
theirs includes the simplifying assumption that the electron density was
uniform in the disk.

Waters {\it et al.} assume $T_{sh} = 0.8T_{\star}$ and calculate
$R_{sh}$ and the density profile ($\rho (r)$) for the stellar disks
using the ``curve of growth'' method (Lamers \& Waters 1984).  For this
method, they assume a disk with an opening angle $\theta$, a disk
density which is proporational to $r^{-n}$, a density at the inner edge
of the shell of $\rho _o$, and a disk radius $r_{sh}$ (the sharp outer
cutoff to the disk).  For their modelling, they have assumed a pole-on
view of the disk, such that the opening angle $\theta$ is deprecated in
their calculations; this is contrary to the both our assumption and the
assumption of Gehrz et al. that the disk is viewed edge-on.  They then
use a combination of optically thin and optically thick free-free
radiation as a comparision to the observed monochromatic infrared excess
(as found from IRAS spectra and using assumed stellar parameters).  By
comparing the shapes of the observed excess and their models, they
obtain a value fo the density parameter $n$, and from the shift required
to match the observed and model curves they obtain $\rho _o$.  These
models are dependent upon the assumed shell radius, such that the
authors run a series of models to find the best-fit combination of $n$
and $r_{sh}$, yielding the values used for comparision in this paper
(notably, their values for $r_{sh}$ are all lower limits).  We do not
compare our values for $n_e$ with the $\rho (r)$ calculated by Waters
{\it et al.} because of the large uncertainty in the mass densities they
derived, the uncertainty in the proper ionization fraction of the
envelope gas, and the dependence upon the exact shell radius.

We present the results of the modelling discussed in this paper, as well
as the results of both Gehrz et al. and Waters et al. in Table~7 for
ease of comparision.  We find that our results are generally consistent
(within factors of a few) with those of both authors.  However, in a
number of the cases, Waters {\it et al.} derive shell radii which are
significantly larger than the radii derived here.  This can be explained
by the assumptions which have gone into our modelling, and the
wavelength regimes of the observations.  We have assumed an isothermal
disk, whereas Waters {\it et al.} did not.  Since their observations
utilize IRAS data, they will be sensitive to the cool parts of the shell
which should exist in the case of a thermal gradient.  These
discrepancies are particularly apparent in the cases of $\phi$~Per and
$\gamma$~Cas, bright sources with accurate 12\micron , 25\micron, and
60\micron \ IRAS detections.

Comparing our results to those of Gehrz {\it et al.}, we find that in
the case of $\zeta$~Tau we predict an electron density which is a
factor of two below that of Gehrz {\it et al.}, and a moderately
larger disk.  In the cases of $\gamma$~Cas and $\phi$~Per, we derive
larger, hotter disks than Gehrz {\it et al.}, but have similar
electron densities.  The simplicity of the models used by Gehrz {\it
et al.} are most likely responsible for this discrepancy.  They
extrapolated curves for both the optically-thin and optically-thick
segments of the free-free emission, then used the intercept point of
these two curves to derive numerical values for radius and temperature
of the shell.  The improved modelling techniques used here should be
better able to limit these same parameters.

In Table~7, we also show calculated total mass of hydrogen in the
emitting disk.  These numbers assume complete ionization of the
hydrogen gas, such that $n_e = n_H$, and uniform density, an
assumption already used in the modelling.  We find that the masses of
the disks are relatively small, ranging from 2.5$\times 10^{-11} \rm
M_{\sun}$ ($o$~Aqr) to 8.1$\times 10^{-9} \rm M_{\sun}$
($\gamma$~Cas).

\section{Discussion}

These spectra present MIR features of Be stars and give some indications
of the range of emission characteristics of them as well.  Continuum
emission is explained by a combination of optically-thick and
optically-thin free-free emission.  Both the continuum and line emission
is likely to arise in a warm stellar envelope.  The peculiar objects
($\beta$~Lyr and MWC~349) display complicated line emission which is
difficult to explain with simple models.  In the case of $\beta$~Lyr, a
relatively simple explanation for the presence of hydrogen and helium is
possible, and the observed [Ne II] emission is easily explained as
emission from a large gas cloud surrounding the binary system.  In the
case of MWC~349, the observed features are consistent with previous
observations, and the observed forbidden emission implies a large volume
of low-density ionized gas.

\acknowledgements

The authors would like to thank the staff at Palomar Observatory for
their assistance.  We would also like to thank S. Hony for sharing ISO
data and results prior to publication.  SAR acknowledges L.B.F.M. Waters
and M. Simon for helpful discussions.  This research was partly
supported by NASA Contract 960803.

\newpage

\begin{table}
\begin{center}
\caption{Observations of Be Stars}
\begin{tabular}{lllll}\hline\hline
Object & Date & Integration & Spectral & IRAS \\
& Observed & Time (s) & Type & 12\micron \ Flux (Jy) \\\tableline
$\psi$ Per & Aug. 29, 1998 & 668 & B5Ve & \\
$\eta$ Tau & Aug. 29, 1998 & 653 & B7III& 4.32\\
$\zeta$ Tau & Aug. 28, 1998 & 481 & B2IV & 8.30\\
48 Per & Aug. 28, 1998 & 661 & B3Ve & 3.80\\
$\phi$ Per & Aug. 27, 1998 & 741 & B2Vpe &\\
$\gamma$ Cas & Aug. 27, 1998 & 240 & B0IVe & 18.83\\
$o$ Aqr & Aug. 27, 1998 & 586 & B7IVe & 1.18\\
$\beta$ Lyr & July 5, 1998 & 930 & B7Ve+ & 5.01\\
& July 13, 1998 & & &\\
EW Lac & July 12, 1998 & 661 & B3IVpe & \\
$\kappa$ Dra & July 5, 1998 & 1094 & B6IIIpe & 4.12 \\
MWC 349 & July 12, 1998 & 481 & Bpe & 179.0 \\ 
        & Aug. 27, 1998 & & & \\\hline
\end{tabular}
\end{center}
\end{table}

\makeatletter
\def\jnl@aj{AJ}
\ifx\revtex@jnl\jnl@aj\let=tablebreak=\nl\fi
\makeatother

\renewcommand{\arraystretch}{0.75}

\begin{deluxetable}{llllc}
\tablewidth{0pc}
\tablenum{2A}
\tablecaption{Data on observed lines -- $\psi$ Per}
\tablehead{
\colhead{Transition}                    &
\colhead{$\lambda _{theory}$}           &
\colhead{$\lambda _{obs}$}              &
\colhead{$\Delta\lambda$}               &
\colhead{$F_{\nu}$}\\
\colhead{}		&
\colhead{ (\micron )}		&
\colhead{ (\micron )}              &
\colhead{($10^{-3}$\micron )}           &
\colhead{($10^{-19}$ W cm$^{-2}$ \micron $^{-1}$)}
}
\startdata
H 14$\rightarrow$9 & 12.587 &
12.582 $\pm$ 0.0064 & 13.40 $\pm$ 5.833 &   6.73 $\pm$  3.96 \cr
H 19$\rightarrow$10 & 12.611 &
12.613 $\pm$ 0.0024 &  4.06 $\pm$ 2.622 &   3.04 $\pm$  2.93 \cr
H 11$\rightarrow$8 & 12.387 &
12.391 $\pm$ 0.0047 &  3.55 $\pm$ 4.034 &   2.81 $\pm$  4.21 \cr
H 7$\rightarrow$6 & 12.372 &
12.371 $\pm$ 0.0016 &  7.18 $\pm$ 1.236 &   10.60 $\pm$ 2.43 \cr
H 21$\rightarrow$10 & 11.792 &
11.787 $\pm$ 0.0046 &  8.93 $\pm$ 4.136 &   4.14 $\pm$  2.56 \cr
H 15$\rightarrow$9 & 11.540 &
11.540 $\pm$ 0.0042 & 13.10 $\pm$ 4.098 &   7.83 $\pm$  3.09 \cr
H 9$\rightarrow$7 & 11.309 &
11.308 $\pm$ 0.0013 & 82.50 $\pm$ 1.216 &   9.86 $\pm$  1.91 \cr
H 24$\rightarrow$10 & 11.033 &
11.033 $\pm$ 0.0049 &  6.91 $\pm$ 4.302 &   2.82 $\pm$  2.24 \cr
H 25$\rightarrow$10 & 10.855 &
10.854 $\pm$ 0.0041 &  5.78 $\pm$ 3.545 &   2.55 $\pm$  1.89 \cr
H 16$\rightarrow$9 & 10.804 &
10.803 $\pm$ 0.0018 &  7.52 $\pm$ 1.640 &   7.09 $\pm$  2.03 \cr
H 26$\rightarrow$10 & 10.701 &
10.695 $\pm$ 0.0053 &  8.00 $\pm$ 1.414 &   3.03 $\pm$  1.67 \cr
H 12$\rightarrow$8 & 10.504 &
10.502 $\pm$ 0.0021 &  8.11 $\pm$ 1.920 &   5.97 $\pm$  1.90 \cr
H 17$\rightarrow$9 & 10.261 &
10.256 $\pm$ 0.0032 & 12.90 $\pm$ 2.921 &   7.87 $\pm$  2.40 \cr
H 30$\rightarrow$10 & 10.258 & \multicolumn{3}{c}{Blended with 17$\rightarrow$9} \cr
H 18$\rightarrow$9 & 9.847 &
 9.842 $\pm$ 0.0035 &  8.42 $\pm$ 3.154 &   6.40 $\pm$  3.09 \cr
H 19$\rightarrow$9 & 9.522 &
 9.518 $\pm$ 0.0043 &  6.66 $\pm$ 3.782 &   3.31 $\pm$  2.59 \cr
H 13$\rightarrow$8 & 9.392 &
 9.390 $\pm$ 0.0013 &  6.11 $\pm$ 1.192 &   9.35 $\pm$  2.29 \cr
H 20$\rightarrow$9 & 9.261 &
 9.258 $\pm$ 0.0051 &  8.63 $\pm$ 4.691 &   4.39 $\pm$  3.10 \cr
H 21$\rightarrow$9 & 9.047 &
 9.044 $\pm$ 0.0046 &  4.35 $\pm$ 3.810 &   2.36 $\pm$  2.57 \cr
H 10$\rightarrow$7 & 8.760 &
 8.758 $\pm$ 0.0012 &  4.97 $\pm$ 1.373 &   7.77 $\pm$  2.78 \cr
H 14$\rightarrow$8 & 8.665  &
 8.663 $\pm$ 0.0022 &  8.41 $\pm$ 1.997 &   10.34 $\pm$ 3.23 \cr
H 24$\rightarrow$9 & 8.594 &
 8.594 $\pm$ 0.0033 &  4.61 $\pm$ 3.466 &   3.75 $\pm$  3.47 \cr
\enddata
\end{deluxetable}

\makeatletter
\def\jnl@aj{AJ}
\ifx\revtex@jnl\jnl@aj\let=tablebreak=\nl\fi
\makeatother

\renewcommand{\arraystretch}{0.75}

\begin{deluxetable}{llllc}
\tablenum{2B}
\tablewidth{0pc}
\tablecaption{Data on observed lines -- $\eta$ Tau}
\tablehead{
\colhead{Transition}                    &
\colhead{$\lambda _{theory}$}           &
\colhead{$\lambda _{obs}$}              &
\colhead{$\Delta\lambda$}               &
\colhead{$F_{\nu}$}\\
\colhead{}              &
\colhead{ (\micron )}              &
\colhead{ (\micron )}              &
\colhead{($10^{-3}$\micron )}           &
\colhead{($10^{-19}$ W cm$^{-2}$ \micron $^{-1}$)}
}
\startdata
H 14$\rightarrow$9 & 12.611 &
12.586 $\pm$ 0.0032 &  9.14 $\pm$ 3.075 &   5.07 $\pm$  2.2.8 \cr
H 11$\rightarrow$8 & 12.387 &
12.393 $\pm$ 0.0016 &  4.21 $\pm$ 1.782 &   2.80 $\pm$  1.74 \cr
H 7$\rightarrow$6 & 12.372 &
12.375 $\pm$ 0.0010 &  5.68 $\pm$ 1.392 &   6.76 $\pm$  2.07 \cr
H 15$\rightarrow$9 & 11.540 &
11.538 $\pm$ 0.0026 &  8.84 $\pm$ 2.376 &   5.27 $\pm$  1.86 \cr
H 9$\rightarrow$7 & 11.309 &
11.308 $\pm$ 0.0009 &  7.41 $\pm$ 0.870 &   8.67 $\pm$  1.33 \cr
H 23$\rightarrow$10 & 11.243 &
11.245 $\pm$ 0.0070 &                  \nodata  &    \nodata  \cr
H 24$\rightarrow$10 & 11.033 &
11.044 $\pm$ 0.0093 &                  \nodata  &    \nodata  \cr
H 25$\rightarrow$10 & 10.855 &
10.859 $\pm$ 0.0042 &  6.02 $\pm$ 5.527 &   1.75 $\pm$  2.17 \cr
H 16$\rightarrow$9 & 10.804 &
10.802 $\pm$ 0.0018 &  7.74 $\pm$ 1.621 &   6.50 $\pm$  1.79 \cr
H 12$\rightarrow$8 & 10.504 &
10.501 $\pm$ 0.0016 &  7.65 $\pm$ 1.483 &   5.61 $\pm$  1.43 \cr
H 17$\rightarrow$9 & 10.261 &
10.257 $\pm$ 0.0028 & 10.00 $\pm$ 2.586 &   4.97 $\pm$  1.73 \cr
H 18$\rightarrow$9 & 9.847 &
 9.844 $\pm$ 0.0026 &  6.72 $\pm$ 2.346 &   4.43 $\pm$  2.05 \cr
H 19$\rightarrow$9 & 9.522 &
 9.516 $\pm$ 0.0026 &  5.23 $\pm$ 2.019 &   3.10 $\pm$  1.65 \cr
H 13$\rightarrow$8 & 9.392 &
 9.389 $\pm$ 0.0019 &  3.14 $\pm$ 2.605 &   2.62 $\pm$  2.89 \cr
H 10$\rightarrow$7 & 8.760 &
 8.758 $\pm$ 0.0016 &  6.42 $\pm$ 1.429 &   9.55 $\pm$  2.80 \cr
H 14$\rightarrow$8 & 8.665 &
 8.663 $\pm$ 0.0023 &  7.19 $\pm$ 2.113 &   8.19 $\pm$  3.22 \cr
H 24$\rightarrow$9 & 8.594 &
 8.595 $\pm$ 0.0084 &                  \nodata  &    \nodata  \cr
H 15$\rightarrow$8 & 8.155 &
 8.164 $\pm$ 0.0017 &  5.57 $\pm$ 1.506 &   7.33 $\pm$  2.65 \cr
H 29$\rightarrow$9 & 8.173 & \multicolumn{3}{c}{Blended with 15$\rightarrow$8}
\enddata
\end{deluxetable}

\makeatletter
\def\jnl@aj{AJ}
\ifx\revtex@jnl\jnl@aj\let=tablebreak=\nl\fi
\makeatother

\renewcommand{\arraystretch}{0.75}

\begin{deluxetable}{llllc}
\tablewidth{0pc}
\tablenum{2C}
\tablecaption{Data on observed lines -- $\zeta$ Tau}
\tablehead{
\colhead{Transition}                    &
\colhead{$\lambda _{theory}$}           &
\colhead{$\lambda _{obs}$}              &
\colhead{$\Delta\lambda$}               &
\colhead{$F_{\nu}$}\\
\colhead{}              &
\colhead{ (\micron ) }             &
\colhead{ (\micron )}              &
\colhead{($10^{-3}$\micron )}           &
\colhead{($10^{-19}$ W cm$^{-2}$ \micron $^{-1}$)}
}
\startdata
H 18$\rightarrow$10 & 13.188 &
13.191 $\pm$ 0.0025 &  7.09 $\pm$ 2.144 &   4.41 $\pm$  1.76 \cr
H 19$\rightarrow$10 & 12.611 &
12.611 $\pm$ 0.0036 &  8.54 $\pm$ 3.164 &   5.16 $\pm$  2.27 \cr
H 14$\rightarrow$9 & 12.587 &
12.586 $\pm$ 0.0032 &  6.88 $\pm$ 3.412 &   4.23 $\pm$  2.42 \cr
H 11$\rightarrow$8 & 12.387 &
12.395 $\pm$ 0.0062 &  3.80 $\pm$ 9.182 &   0.91 $\pm$  2.40 \cr
H 7$\rightarrow$6 & 12.372 &
12.375 $\pm$ 0.0039 & 11.60 $\pm$ 3.078 &   8.93 $\pm$  2.89 \cr
H 20$\rightarrow$10 & 12.157 &
12.160 $\pm$ 0.0027 &  7.94 $\pm$ 2.588 &   4.52 $\pm$  1.94 \cr
H 21$\rightarrow$10 & 11.792 &
11.796 $\pm$ 0.0030 & 10.40 $\pm$ 2.794 &   5.05 $\pm$  1.76 \cr
H 15$\rightarrow$9 & 11.540 &
11.542 $\pm$ 0.0025 & 10.00 $\pm$ 2.219 &   6.20 $\pm$  1.83 \cr
H 22$\rightarrow$10 & 11.492 &
11.493 $\pm$ 0.0040 &  8.41 $\pm$ 3.562 &   2.93 $\pm$  1.67 \cr
H 9$\rightarrow$7 & 11.309 &
11.310 $\pm$ 0.0022 &  8.51 $\pm$ 2.032 &   5.46 $\pm$  1.73 \cr
H 23$\rightarrow$10 & 11.243 &
11.243 $\pm$ 0.0032 &  7.54 $\pm$ 2.775 &   3.53 $\pm$  1.72 \cr
H 24$\rightarrow$10 & 11.033 &
11.037 $\pm$ 0.0037 &  6.13 $\pm$ 4.111 &   2.56 $\pm$  2.31 \cr
H 25$\rightarrow$10 & 10.855 &
10.860 $\pm$ 0.0037 &  7.93 $\pm$ 2.941 &   3.95 $\pm$  1.93 \cr
H 16$\rightarrow$9 & 10.804 &
10.804 $\pm$ 0.0016 &  9.71 $\pm$ 1.481 &   10.54 $\pm$ 2.12 \cr
H 26$\rightarrow$10 & 10.701 &
10.696 $\pm$ 0.0036 &  4.66 $\pm$ 4.525 &   2.08 $\pm$  2.53 \cr
H 27$\rightarrow$10 & 10.567 &
10.573 $\pm$ 0.0058 &  4.18 $\pm$ 5.796 &   0.88 $\pm$  1.57 \cr
H 12$\rightarrow$8 & 10.504 &
10.507 $\pm$ 0.0039 &  6.73 $\pm$ 3.551 &   3.74 $\pm$  2.41 \cr
H 28$\rightarrow$10 & 10.451 &
10.458 $\pm$ 0.0080 &  6.84 $\pm$ 7.271 &   1.91 $\pm$  2.48 \cr
H 29$\rightarrow$10 & 10.348 &
10.353 $\pm$ 0.0113 &                  \nodata  &    \nodata  \cr
H 17$\rightarrow$9 & 10.261 &
10.264 $\pm$ 0.0016 &  7.96 $\pm$ 1.505 &   8.71 $\pm$  2.18 \cr
H 30$\rightarrow$10 & 10.258 & \multicolumn{3}{c}{Blended with 17$\rightarrow$9}\cr
H 31$\rightarrow$10 & 10.177 &
10.178 $\pm$ 0.0024 &  3.84 $\pm$ 2.404 &   1.89 $\pm$  1.65 \cr
H 18$\rightarrow$9 & 9.847 &
 9.849 $\pm$ 0.0016 &  7.23 $\pm$ 1.451 &   9.15 $\pm$  2.39 \cr
H 19$\rightarrow$9 & 9.522 &
 9.527 $\pm$ 0.0010 &                  \nodata  &    \nodata  \cr
H 13$\rightarrow$8 & 9.392 &
 9.395 $\pm$ 0.0019 &  7.21 $\pm$ 1.708 &   9.64 $\pm$  3.01 \cr
H 20$\rightarrow$9 & 9.261 &
 9.266 $\pm$ 0.0018 &  5.88 $\pm$ 1.719 &   7.34 $\pm$  2.84 \cr
H 21$\rightarrow$9 & 9.047 &
 9.049 $\pm$ 0.0036 &  6.44 $\pm$ 3.288 &   5.61 $\pm$  3.65 \cr
H 22$\rightarrow$9 & 8.870 &
 8.875 $\pm$ 0.0035 &  5.78 $\pm$ 2.963 &   4.66 $\pm$  3.22 \cr
H 10$\rightarrow$7 & 8.760 &
 8.761 $\pm$ 0.0024 &  5.72 $\pm$ 1.965 &   5.45 $\pm$  3.03 \cr
H 23$\rightarrow$9 & 8.721 &
 8.723 $\pm$ 0.0025 &  6.64 $\pm$ 2.232 &   7.86 $\pm$  3.46 \cr
H 14$\rightarrow$8 & 8.665  &
 8.667 $\pm$ 0.0024 &  6.96 $\pm$ 2.207 &   8.30 $\pm$  3.48 \cr
H 24$\rightarrow$9 & 8.594 &
 8.596 $\pm$ 0.0031 &  5.84 $\pm$ 2.774 &   6.67 $\pm$  4.02 \cr
H 25$\rightarrow$9 & 8.485 &
 8.488 $\pm$ 0.0021 &  2.75 $\pm$ 2.296 &   3.02 $\pm$  3.51 \cr
H 26$\rightarrow$9 & 8.391 &
 8.392 $\pm$ 0.0033 &  4.99 $\pm$ 2.427 &   4.60 $\pm$  3.26 \cr
H 27$\rightarrow$9 & 8.309 &
 8.306 $\pm$ 0.0036 &  4.37 $\pm$ 3.477 &   3.55 $\pm$  3.58 \cr
H 28$\rightarrow$9 & 8.236 &
 8.239 $\pm$ 0.0011 &                  \nodata  &    \nodata  \cr
H 15$\rightarrow$8 & 8.155 &
 8.156 $\pm$ 0.0009 &  4.85 $\pm$ 1.035 &   11.02 $\pm$  3.38 \cr
H 29$\rightarrow$9 & 8.173& \multicolumn{3}{c}{Blended with 15$\rightarrow$8} \cr
H 30$\rightarrow$9 & 8.155 &
 8.103 $\pm$ 0.0023 &  6.33 $\pm$ 2.102 &   8.12 $\pm$  3.74 \cr
\enddata
\end{deluxetable}

\makeatletter
\def\jnl@aj{AJ}
\ifx\revtex@jnl\jnl@aj\let=tablebreak=\nl\fi
\makeatother

\renewcommand{\arraystretch}{0.75}

\begin{deluxetable}{llllc}
\tablenum{2D}
\tablewidth{0pc}
\tablecaption{Data on observed lines -- 48 Per}
\tablehead{
\colhead{Transition}                    &
\colhead{$\lambda _{theory}$}           &
\colhead{$\lambda _{obs}$}              &
\colhead{$\Delta\lambda$}               &
\colhead{$F_{\nu}$}\\
\colhead{}              &
\colhead{ (\micron )}              &
\colhead{ (\micron )}              &
\colhead{($10^{-3}$\micron )}           &
\colhead{($10^{-19}$ W cm$^{-2}$ \micron $^{-1}$)}
}
\startdata
H 18$\rightarrow$10 & 13.188 &
13.188 $\pm$ 0.0065 &  6.14 $\pm$ 4.530 &   3.04 $\pm$  2.95 \cr
H 19$\rightarrow$10 & 12.611 & \multicolumn{3}{c}{Blended with H 14$\rightarrow$9}\cr
H 14$\rightarrow$9 & 12.587 &
12.587 $\pm$ 0.0028 &  9.00 $\pm$ 2.685 &   6.27 $\pm$  2.42 \cr
H 11$\rightarrow$8 & 12.387 & \multicolumn{3}{c}{Blended with H 7$\rightarrow$6}\cr
H 7$\rightarrow$6 & 12.372 &
12.379 $\pm$ 0.0019 &  8.33 $\pm$ 1.594 &   7.72 $\pm$  2.01 \cr
H 20$\rightarrow$10 & 12.157 &
12.159 $\pm$ 0.0033 &  5.90 $\pm$ 3.344 &   2.30 $\pm$  1.66 \cr
H 21$\rightarrow$10 & 11.792 &
11.792 $\pm$ 0.0061 & 10.10 $\pm$ 5.602 &   2.66 $\pm$  1.92 \cr
H 15$\rightarrow$9 & 11.540 &
11.537 $\pm$ 0.0027 &  8.52 $\pm$ 2.482 &   4.18 $\pm$  1.60 \cr
H 9$\rightarrow$7 & 11.309 &
11.306 $\pm$ 0.0011 &  6.41 $\pm$ 0.889 &   6.00 $\pm$  1.12 \cr
H 23$\rightarrow$10 & 11.243 &
11.236 $\pm$ 0.0062 &  7.72 $\pm$ 5.487 &   1.69 $\pm$  1.60 \cr
H 24$\rightarrow$10 & 11.033 &
11.028 $\pm$ 0.0044 &  4.16 $\pm$ 8.134 &   0.95 $\pm$  2.39 \cr
H 16$\rightarrow$9 & 10.804 &
10.802 $\pm$ 0.0019 &  7.31 $\pm$ 1.777 &   4.82 $\pm$  1.56 \cr
H 27$\rightarrow$10 & 10.567 &
10.566 $\pm$ 0.0197 &                  \nodata  &    \nodata  \cr
H 12$\rightarrow$8 & 10.504 &
10.501 $\pm$ 0.0015 &  6.79 $\pm$ 1.334 &   5.05 $\pm$  1.35 \cr
H 28$\rightarrow$10 & 10.451 &
10.446 $\pm$ 0.0226 &                  \nodata  &    \nodata  \cr
H 17$\rightarrow$9 & 10.261 &
10.257 $\pm$ 0.0013 &  7.61 $\pm$ 1.179 &   6.69 $\pm$  1.37 \cr
H 18$\rightarrow$9 & 9.847 &
 9.846 $\pm$ 0.0021 &  6.19 $\pm$ 1.812 &   4.92 $\pm$  1.94 \cr
H 19$\rightarrow$9 & 9.522 &
 9.517 $\pm$ 0.0019 &  4.36 $\pm$ 1.390 &   3.65 $\pm$  1.54 \cr
H 13$\rightarrow$8 & 9.392 &
 9.388 $\pm$ 0.0011 &  5.92 $\pm$ 0.913 &   8.08 $\pm$  1.64 \cr
H 20$\rightarrow$9 & 9.261 &
 9.260 $\pm$ 0.0049 &  9.45 $\pm$ 4.608 &   4.07 $\pm$  2.60 \cr
H 21$\rightarrow$9 & 9.047 &
 9.044 $\pm$ 0.0046 &  6.15 $\pm$ 4.097 &   3.14 $\pm$  2.68 \cr
H 22$\rightarrow$9 & 8.870 &
 8.870 $\pm$ 0.0051 &  2.58 $\pm$ 5.069 &   1.33 $\pm$  2.77 \cr
H 10$\rightarrow$7 & 8.760 &
 8.757 $\pm$ 0.0009 &  5.82 $\pm$ 0.756 &   9.98 $\pm$  1.70 \cr
H 23$\rightarrow$9 & 8.721 &
 8.718 $\pm$ 0.0059 &  5.76 $\pm$ 5.464 &   2.14 $\pm$  2.60 \cr
H 14$\rightarrow$8 & 8.665  &
 8.662 $\pm$ 0.0015 &  5.68 $\pm$ 1.286 &   7.55 $\pm$  2.27 \cr
\enddata
\end{deluxetable}

\makeatletter
\def\jnl@aj{AJ}
\ifx\revtex@jnl\jnl@aj\let=tablebreak=\nl\fi
\makeatother

\renewcommand{\arraystretch}{0.75}

\begin{deluxetable}{llllc}
\tablenum{2E}
\tablewidth{0pc}
\tablecaption{Data on observed lines -- $\phi$ Per}
\tablehead{
\colhead{Transition}                    &
\colhead{$\lambda _{theory}$}           &
\colhead{$\lambda _{obs}$}              &
\colhead{$\Delta\lambda$}               &
\colhead{$F_{\nu}$}\\
\colhead{}              &
\colhead{ (\micron )}              &
\colhead{ (\micron )}              &
\colhead{($10^{-3}$\micron )}           &
\colhead{($10^{-19}$ W cm$^{-2}$ \micron $^{-1}$)}
}
\startdata
H 18$\rightarrow$10 & 13.188 &
13.185 $\pm$ 0.0059 & 13.10 $\pm$ 5.435 &   4.23 $\pm$  2.29 \cr
H 19$\rightarrow$10 & 12.611 &
12.615 $\pm$ 0.0009 &  3.84 $\pm$ 1.099 &   2.53 $\pm$  1.21 \cr
H 14$\rightarrow$9 & 12.587 &
12.588 $\pm$ 0.0024 &  6.90 $\pm$ 1.996 &   2.91 $\pm$  1.11 \cr
H 11$\rightarrow$8 & 12.387 &\multicolumn{3}{c}{Blended with 7$\rightarrow$6} \cr
H 7$\rightarrow$6 & 12.372 &
12.364 $\pm$ 0.0040 & 12.49 $\pm$ 3.691 &   4.41 $\pm$  1.71 \cr
H 20$\rightarrow$10 & 12.157 &
12.155 $\pm$ 0.0048 & 12.20 $\pm$ 4.486 &   2.88 $\pm$  1.38 \cr
H 15$\rightarrow$9 & 11.540 &
11.536 $\pm$ 0.0020 &  7.88 $\pm$ 1.774 &   3.29 $\pm$  0.97 \cr
H 9$\rightarrow$7 & 11.309 &
11.305 $\pm$ 0.0024 & 69.90 $\pm$ 2.115 &   2.38 $\pm$  0.95 \cr
H 23$\rightarrow$10 & 11.243 &
11.238 $\pm$ 0.0039 &  6.00 $\pm$ 3.846 &   1.33 $\pm$  1.04 \cr
H 24$\rightarrow$10 & 11.033 &
11.033 $\pm$ 0.0052 &  9.45 $\pm$ 4.707 &   2.20 $\pm$  1.47 \cr
H 16$\rightarrow$9 & 10.804 &
10.801 $\pm$ 0.0019 &  7.68 $\pm$ 1.812 &   4.24 $\pm$  1.35 \cr
H 26$\rightarrow$10 & 10.701 &
10.698 $\pm$ 0.0036 &  5.10 $\pm$ 2.802 &   1.63 $\pm$  1.17 \cr
H 27$\rightarrow$10 & 10.567 &
10.577 $\pm$ 0.0058 &  6.70 $\pm$ 4.402 &   1.17 $\pm$  1.05 \cr
H 12$\rightarrow$8 & 10.504 &
10.502 $\pm$ 0.0023 &  6.04 $\pm$ 2.163 &   2.55 $\pm$  1.27 \cr
H 28$\rightarrow$10 & 10.451 &
10.452 $\pm$ 0.0067 & 4.690 $\pm$ 6.250 &   7.24 $\pm$  1.22 \cr
H 17$\rightarrow$9 & 10.261 &
10.258 $\pm$ 0.0018 &  7.51 $\pm$ 1.642 &   4.47 $\pm$  1.29 \cr
H 30$\rightarrow$10 & 10.258 & \multicolumn{3}{c}{Blended with 17$\rightarrow$9} \cr
H 18$\rightarrow$9 & 9.847 &
 9.845 $\pm$ 0.0017 &  5.29 $\pm$ 1.330 &   5.25 $\pm$  1.56 \cr
H 19$\rightarrow$9 & 9.522 &
 9.518 $\pm$ 0.0038 &  5.59 $\pm$ 3.424 &   2.92 $\pm$  2.07 \cr
H 13$\rightarrow$8 & 9.392 &
 9.391 $\pm$ 0.0017 &  5.27 $\pm$ 1.597 &   4.55 $\pm$  1.79 \cr
H 20$\rightarrow$9 & 9.261 &
 9.257 $\pm$ 0.0026 &  5.67 $\pm$ 2.591 &   3.49 $\pm$  2.02 \cr
H 21$\rightarrow$9 & 9.047 &
 9.047 $\pm$ 0.0022 &  5.20 $\pm$ 2.006 &   3.49 $\pm$  1.76 \cr
H 22$\rightarrow$9 & 8.870 &
 8.871 $\pm$ 0.0073 & 10.41 $\pm$ 7.766 &   6.53 $\pm$  5.99 \cr
H 10$\rightarrow$7 & 8.760 &
 8.758 $\pm$ 0.0043 &  5.43 $\pm$ 5.002 &   3.40 $\pm$  4.26 \cr
H 29$\rightarrow$9 & 8.173& \multicolumn{3}{c}{Blended with 15$\rightarrow$8} \cr
H 15$\rightarrow$8 & 8.155 &
 8.152 $\pm$ 0.0014 & 3.81$\pm$1.021 & 7.0299$\pm$2.5343 \cr
\enddata
\end{deluxetable}

\makeatletter
\def\jnl@aj{AJ}
\ifx\revtex@jnl\jnl@aj\let=tablebreak=\nl\fi
\makeatother

\renewcommand{\arraystretch}{0.75}

\begin{deluxetable}{lllll}
\tablenum{2}
\tablewidth{0pc}
\tablecaption{Data on observed lines}
\tablehead{\colhead{}				&
\colhead{}				&
\multicolumn{3}{c}{$\gamma$~Cas, Aug 1998}	\\
\colhead{Transition}			&
\colhead{$\lambda _{theory}$}		&
\colhead{$\lambda _{obs}$}              &
\colhead{$\Delta\lambda$}               &
\colhead{$F_{\nu}$}
}
\startdata
H 7$\rightarrow$6 & 12.372 &
12.369 $\pm$ 0.0031 &  9.53 $\pm$ 2.643 &   1.09 $\pm$  0.363 \cr
H 9$\rightarrow$7 & 11.309 &
11.307 $\pm$ 0.0015 &  6.57 $\pm$ 1.391 &   1.20 $\pm$  0.328 \cr
H 10$\rightarrow$7 & 8.760 &
 8.759 $\pm$ 0.0031 &  6.13 $\pm$ 2.804 &   2.19 $\pm$  1.328 \cr
H 11$\rightarrow$8 & 12.387 &
12.394 $\pm$ 0.0065 &  7.71 $\pm$ 6.266 &   0.45 $\pm$  0.433 \cr
H 12$\rightarrow$8 & 10.504 &
10.504 $\pm$ 0.0018 &  6.85 $\pm$ 1.670 &   1.25 $\pm$  0.400 \cr
H 13$\rightarrow$8 & 9.392 &
 9.393 $\pm$ 0.0015 &  6.85 $\pm$ 1.386 &   1.79 $\pm$  0.474 \cr
H 14$\rightarrow$8 & 8.665  &
 8.666 $\pm$ 0.0028 &  7.69 $\pm$ 2.543 &   2.33 $\pm$  1.174 \cr
H 14$\rightarrow$9 & 12.587 &
12.589 $\pm$ 0.0022 &  6.37 $\pm$ 1.716 &   1.05 $\pm$  0.370 \cr
H 15$\rightarrow$8 & 8.155 &
 8.154 $\pm$ 0.0010 &  4.20 $\pm$ 0.832 &   4.96 $\pm$  1.311 \cr
H 15$\rightarrow$9 & 11.540 &
11.540 $\pm$ 0.0017 &  8.03 $\pm$ 1.555 &   1.11 $\pm$  0.283 \cr
H 16$\rightarrow$9 & 10.804 &
10.803 $\pm$ 0.0015 &  6.39 $\pm$ 1.314 &   1.41 $\pm$  0.383 \cr
H 17$\rightarrow$9 & 10.261 &
10.260 $\pm$ 0.0018 &  7.62 $\pm$ 1.620 &   1.28 $\pm$  0.361 \cr
H 18$\rightarrow$9 & 9.847 &
 9.846 $\pm$ 0.0016 &  6.38 $\pm$ 1.349 &   1.74 $\pm$  0.494 \cr
H 18$\rightarrow$10 & 13.188 &
13.197 $\pm$ 0.0034 &  6.70 $\pm$ 2.798 &   0.74 $\pm$  0.410 \cr
H 19$\rightarrow$9 & 9.522 &
 9.522 $\pm$ 0.0023 &  2.70 $\pm$ 2.071 &   0.74 $\pm$  0.721 \cr
H 19$\rightarrow$10 & 12.611 &
12.615 $\pm$ 0.0021 &  5.63 $\pm$ 2.998 &   0.87 $\pm$  0.615 \cr
H 20$\rightarrow$9 & 9.261 &
 9.262 $\pm$ 0.0035 &  9.17 $\pm$ 3.226 &   0.88 $\pm$  0.433 \cr
H 20$\rightarrow$10 & 12.157 &
12.157 $\pm$ 0.0044 &  9.53 $\pm$ 4.018 &   0.46 $\pm$  0.259 \cr
H 21$\rightarrow$9 & 9.047 &
 9.048 $\pm$ 0.0017 &  2.27 $\pm$ 2.738 &   2.17 $\pm$  5.413 \cr
H 21$\rightarrow$10 & 11.792 &
11.793 $\pm$ 0.0043 &  7.66 $\pm$ 3.738 &   0.52 $\pm$  0.334 \cr
H 22$\rightarrow$9 & 8.870 &
 8.873 $\pm$ 0.0057 &  6.64 $\pm$ 5.096 &   1.34 $\pm$  1.340 \cr
H 22$\rightarrow$10 & 11.492 &
11.488 $\pm$ 0.0032 &  4.14 $\pm$ 4.405 &   0.36 $\pm$  0.537 \cr
H 23$\rightarrow$9 & 8.721 &
 8.726 $\pm$ 0.0172 &                  \nodata  &    \nodata  \cr
H 23$\rightarrow$10 & 11.243 &
11.241 $\pm$ 0.0054 &  7.70 $\pm$ 4.900 &   0.39 $\pm$  0.334 \cr
H 24$\rightarrow$9 & 8.594 &
 8.594 $\pm$ 0.0021 &  2.10 $\pm$ 2.184 &   2.02 $\pm$  3.400 \cr
H 24$\rightarrow$10 & 11.033 &
11.034 $\pm$ 0.0061 &  8.01 $\pm$ 5.466 &   0.43 $\pm$  0.383 \cr
H 25$\rightarrow$9 & 8.485 &
 8.489 $\pm$ 0.0050 &  6.74 $\pm$ 4.387 &   1.43 $\pm$  1.237 \cr
H 25$\rightarrow$10 & 10.855 &
10.849 $\pm$ 0.0099 &                  \nodata  &    \nodata  \cr
H 26$\rightarrow$9 & 8.391 &
 8.390 $\pm$ 0.0036 &  4.28 $\pm$ 3.611 &   1.61 $\pm$  1.737 \cr
H 26$\rightarrow$10 & 10.701 &
10.694 $\pm$ 0.0096 &                  \nodata  &    \nodata  \cr
H 27$\rightarrow$9 & 8.309 &
 8.307 $\pm$ 0.0053 &                  \nodata  &    \nodata  \cr
H 27$\rightarrow$10 & 10.567 &
10.580 $\pm$ 0.0091 &                  \nodata  &    \nodata  \cr
H 28$\rightarrow$9 & 8.236 &
 8.240 $\pm$ 0.0018 &  3.65 $\pm$ 1.244 &   3.58 $\pm$  1.550 \cr
H 28$\rightarrow$10 & 10.451 &
10.453 $\pm$ 0.0248 &                  \nodata  &    \nodata  \cr
H 29$\rightarrow$9 & 8.173& \multicolumn{3}{c}{Blended with 15$\rightarrow$8} \cr
H 29$\rightarrow$10 & 10.348 &
10.334 $\pm$ 0.0049 &                  \nodata  &    \nodata  \cr
H 30$\rightarrow$9 & 8.116 &
 8.107 $\pm$ 0.0040 &  6.80 $\pm$ 3.654 &   1.71 $\pm$  1.218 \cr
H 30$\rightarrow$10 & 10.258 & \multicolumn{3}{c}{Blended with 17$\rightarrow$9} \cr
H 31$\rightarrow$10 & 10.177 &
10.181 $\pm$ 0.0074 &  6.71 $\pm$ 6.115 &   0.28 $\pm$  0.332 \cr
\enddata
\end{deluxetable}

\makeatletter
\def\jnl@aj{AJ}
\ifx\revtex@jnl\jnl@aj\let=tablebreak=\nl\fi
\makeatother

\renewcommand{\arraystretch}{0.75}

\begin{deluxetable}{llllc}
\tablenum{2G}
\tablewidth{0pc}
\tablecaption{Data on observed lines -- $o$ Aqr}
\tablehead{
\colhead{Transition}                    &
\colhead{$\lambda _{theory}$}           &
\colhead{$\lambda _{obs}$}              &
\colhead{$\Delta\lambda$}               &
\colhead{$F_{\nu}$}\\
\colhead{}              &
\colhead{ (\micron )}              &
\colhead{ (\micron )}              &
\colhead{($10^{-3}$\micron )}           &
\colhead{($10^{-19}$ W cm$^{-2}$ \micron $^{-1}$)}
}
\startdata
H 19$\rightarrow$10 & 12.611 &
12.618 $\pm$ 0.0162 &  9.24 $\pm$ 13.21 &   2.03 $\pm$  19.08 \cr
H 14$\rightarrow$9 & 12.587 &
12.590 $\pm$ 0.0041 & 10.84 $\pm$ 3.713 &   2.78 $\pm$  1.23 \cr
H 11$\rightarrow$8 & 12.387 &
12.384 $\pm$ 0.0084 & 21.00 $\pm$ 7.290 &   4.24 $\pm$  1.98 \cr
H 7$\rightarrow$6 & 12.372 &
12.370 $\pm$ 0.0027 &  5.44 $\pm$ 1.812 &   1.86 $\pm$  0.83 \cr
H 21$\rightarrow$10 & 11.792 &
11.782 $\pm$ 0.0061 & 12.13 $\pm$ 5.769 &   1.43 $\pm$  0.94 \cr
H 15$\rightarrow$9 & 11.540 &
11.543 $\pm$ 0.0035 &  9.75 $\pm$ 3.229 &   18.11 $\pm$  6.02 \cr
H 9$\rightarrow$7 & 11.309 &
11.307 $\pm$ 0.0020 &  9.35 $\pm$ 1.813 &   2.68 $\pm$  0.69 \cr
H 24$\rightarrow$10 & 11.033 &
11.024 $\pm$ 0.0081 &                  \nodata  &    \nodata  \cr
H 16$\rightarrow$9 & 10.804 &
10.805 $\pm$ 0.0036 &  7.73 $\pm$ 3.211 &   1.33 $\pm$  0.72 \cr
H 12$\rightarrow$8 & 10.504 &
10.502 $\pm$ 0.0013 &  7.26 $\pm$ 1.142 &   2.89 $\pm$  0.60 \cr
H 28$\rightarrow$10 & 10.451 &
10.453 $\pm$ 0.0264 &                  \nodata  &    \nodata  \cr
H 29$\rightarrow$10 & 10.348 &
10.343 $\pm$ 0.0020 &                  \nodata  &    \nodata  \cr
H 17$\rightarrow$9 & 10.261 &
10.260 $\pm$ 0.0015 &  4.80 $\pm$ 1.005 &   1.72 $\pm$  0.48 \cr
H 30$\rightarrow$10 & 10.258 & \multicolumn{3}{c}{Blended with 17$\rightarrow$9} \cr
H 18$\rightarrow$9 & 9.847 &
 9.846 $\pm$ 0.0029 &  4.70 $\pm$ 2.315 &   1.33 $\pm$  0.81 \cr
H 13$\rightarrow$8 & 9.392 &
 9.392 $\pm$ 0.0023 &  8.80 $\pm$ 2.075 &   3.30 $\pm$  1.04 \cr
H 22$\rightarrow$9 & 8.870 &
 8.871 $\pm$ 0.0116 &                  \nodata  &    \nodata  \cr
H 10$\rightarrow$7 & 8.760 &
 8.759 $\pm$ 0.0026 &  7.86 $\pm$ 2.421 &   4.61 $\pm$  1.86 \cr
H 14$\rightarrow$8 & 8.665  &
 8.668 $\pm$ 0.0024 &  5.10 $\pm$ 2.145 &   3.00 $\pm$  1.67 \cr
H 15$\rightarrow$8 & 8.155 &
 8.155 $\pm$ 0.0011 &  4.43 $\pm$ 1.011 &   4.68 $\pm$  1.40 \cr
H 29$\rightarrow$9 & 8.173& \multicolumn{3}{c}{Blended with 15$\rightarrow$8} \cr
\enddata
\end{deluxetable}

\makeatletter
\def\jnl@aj{AJ}
\ifx\revtex@jnl\jnl@aj\let=tablebreak=\nl\fi
\makeatother

\renewcommand{\arraystretch}{0.75}

\begin{deluxetable}{llllc}
\tablenum{2H}
\tablewidth{0pc}
\tablecaption{Data on observed lines -- EW Lac}
\tablehead{
\colhead{Transition}                    &
\colhead{$\lambda _{theory}$}           &
\colhead{$\lambda _{obs}$}              &
\colhead{$\Delta\lambda$}               &
\colhead{$F_{\nu}$}\\
\colhead{}              &
\colhead{ (\micron )}              &
\colhead{ (\micron )}              &
\colhead{($10^{-3}$\micron )}           &
\colhead{($10^{-19}$ W cm$^{-2}$ \micron $^{-1}$)}
}
\startdata
H 19$\rightarrow$10 & 12.611 &
12.614 $\pm$ 0.0076 &  8.70 $\pm$ 6.928 &   1.24 $\pm$  1.16 \cr
H 14$\rightarrow$9 & 12.587 &
12.591 $\pm$ 0.0045 &  8.40 $\pm$ 3.521 &   2.03 $\pm$  1.05 \cr
H 11$\rightarrow$8 & 12.387 & \multicolumn{3}{c}{Blended with H 7$\rightarrow$6} \cr
H 7$\rightarrow$6 & 12.372 &
12.373 $\pm$ 0.0017 & 12.40 $\pm$ 1.549 &   3.73 $\pm$  0.62 \cr
H 15$\rightarrow$9 & 11.540 &
11.543 $\pm$ 0.0026 &  8.20 $\pm$ 2.307 &   0.88 $\pm$  0.70 \cr
H 22$\rightarrow$10 & 11.492 &
11.491 $\pm$ 0.0100 &                  \nodata  &    \nodata  \cr
H 9$\rightarrow$7 & 11.309 &
11.308 $\pm$ 0.0017 &  8.40 $\pm$ 1.575 &   1.66 $\pm$  0.41 \cr
H 16$\rightarrow$9 & 10.804 &
10.805 $\pm$ 0.0023 &  6.60 $\pm$ 1.876 &   0.68 $\pm$  0.27 \cr
H 26$\rightarrow$10 & 10.701 &
10.696 $\pm$ 0.0029 &  4.83 $\pm$ 3.181 &   0.39 $\pm$  0.32 \cr
H 12$\rightarrow$8 & 10.504 &
10.500 $\pm$ 0.0021 &  7.60 $\pm$ 1.871 &   1.09 $\pm$  0.36 \cr
H 17$\rightarrow$9 & 10.261 &
10.263 $\pm$ 0.0027 & 14.00 $\pm$ 2.669 &   1.78 $\pm$  1.16 \cr
H 30$\rightarrow$10 & 10.258 & \multicolumn{3}{c}{Blended with 17$\rightarrow$9} \cr
H 19$\rightarrow$9 & 9.522 &
 9.523 $\pm$ 0.0024 &  5.50 $\pm$ 2.010 &   1.41 $\pm$  1.44 \cr
H 13$\rightarrow$8 & 9.392 &
 9.386 $\pm$ 0.0026 &  5.70 $\pm$ 2.528 &   1.11 $\pm$  1.39 \cr
H 20$\rightarrow$9 & 9.261 &
 9.254 $\pm$ 0.0043 &  3.86 $\pm$ 2.720 &   5.50 $\pm$  4.00 \cr
H 21$\rightarrow$9 & 9.047 &
 9.048 $\pm$ 0.0027 &  2.50 $\pm$ 3.141 &   3.96 $\pm$  5.02 \cr
H 10$\rightarrow$7 & 8.760 &
 8.758 $\pm$ 0.0014 &  6.10 $\pm$ 1.321 &   1.79 $\pm$  1.11 \cr
H 14$\rightarrow$8 & 8.665  &
 8.664 $\pm$ 0.0031 &  6.60 $\pm$ 2.555 &   0.85 $\pm$  0.97 \cr
\enddata
\end{deluxetable}

\makeatletter
\def\jnl@aj{AJ}
\ifx\revtex@jnl\jnl@aj\let=tablebreak=\nl\fi
\makeatother

\renewcommand{\arraystretch}{0.75}

\begin{deluxetable}{llllc}
\tablewidth{0pc}
\tablenum{2I}
\tablecaption{Data on observed lines -- $\kappa$ Dra}
\tablehead{
\colhead{Transition}                    &
\colhead{$\lambda _{theory}$}           &
\colhead{$\lambda _{obs}$}              &
\colhead{$\Delta\lambda$}               &
\colhead{$F_{\nu}$}\\
\colhead{}			   &
\colhead{ (\micron )}              &
\colhead{ (\micron )}              &
\colhead{($10^{-3}$\micron )}           &
\colhead{($10^{-19}$ W cm$^{-2}$ \micron $^{-1}$)}
}
\startdata
H 18$\rightarrow$10 & 13.188 &
13.192 $\pm$ 0.0069 &  6.30 $\pm$ 7.434 &   1.67 $\pm$  2.47 \cr
H 19$\rightarrow$10 & 12.611 &
12.617 $\pm$ 0.0026 &  7.50 $\pm$ 2.327 &   1.34 $\pm$  1.73 \cr
H 14$\rightarrow$9 & 12.587 &
12.590 $\pm$ 0.0041 &  8.40 $\pm$ 3.497 &   2.56 $\pm$  2.30 \cr
H 11$\rightarrow$8 & 12.387 &
12.393 $\pm$ 0.0026 &  3.54 $\pm$ 3.874 &   0.18 $\pm$  2.83 \cr
H 7$\rightarrow$6 & 12.372 &
12.376 $\pm$ 0.0017 &  7.55 $\pm$ 1.727 &   0.35 $\pm$  1.75 \cr
H 20$\rightarrow$10 & 12.157 &
12.155 $\pm$ 0.0046 &  6.93 $\pm$ 3.672 &   1.07 $\pm$  1.73 \cr
H 21$\rightarrow$10 & 11.792 &
11.793 $\pm$ 0.0059 & 12.70 $\pm$ 5.418 &   1.96 $\pm$  2.42 \cr
H 15$\rightarrow$9 & 11.540 &
11.539 $\pm$ 0.0021 & 10.30 $\pm$ 1.897 &   3.03 $\pm$  1.61 \cr
H 22$\rightarrow$10 & 11.492 &
11.491 $\pm$ 0.0055 &  8.00 $\pm$ 4.943 &   0.90 $\pm$  0.75 \cr
H 9$\rightarrow$7 & 11.309 &
11.309 $\pm$ 0.0014 &  7.60 $\pm$ 1.341 &   2.75 $\pm$  1.35 \cr
H 23$\rightarrow$10 & 11.243 &
11.244 $\pm$ 0.0061 &  7.80 $\pm$ 5.390 &   0.87 $\pm$  1.76 \cr
H 25$\rightarrow$10 & 10.855 &
10.847 $\pm$ 0.0042 &  8.01 $\pm$ 12.14 &   1.16 $\pm$  2.33 \cr
H 16$\rightarrow$9 & 10.804 &
10.803 $\pm$ 0.0014 & 79.40 $\pm$ 1.301 &   3.39 $\pm$  1.61 \cr
H 26$\rightarrow$10 & 10.701 &
10.701 $\pm$ 0.0027 &                  \nodata  &    \nodata  \cr
H 12$\rightarrow$8 & 10.504 &
10.504 $\pm$ 0.0017 &  8.00 $\pm$ 1.520 &   3.10 $\pm$  1.70 \cr
H 17$\rightarrow$9 & 10.261 &
10.263 $\pm$ 0.0017 &  8.50 $\pm$ 1.528 &   3.33 $\pm$  1.74 \cr
H 18$\rightarrow$9 & 9.847 &
 9.845 $\pm$ 0.0022 &  6.92 $\pm$ 1.979 &   2.71 $\pm$  1.03 \cr
H 19$\rightarrow$9 & 9.522 &
 9.521 $\pm$ 0.0023 &  8.57 $\pm$ 2.067 &   3.49 $\pm$  1.11 \cr
H 13$\rightarrow$8 & 9.392 &
 9.390 $\pm$ 0.0014 &  6.75 $\pm$ 1.202 &   3.84 $\pm$  0.93 \cr
H 20$\rightarrow$9 & 9.261 &
 9.260 $\pm$ 0.0034 &  5.20 $\pm$ 2.654 &   12.16 $\pm$  6.23 \cr
H 21$\rightarrow$9 & 9.047 &
 9.047 $\pm$ 0.0033 &  7.82 $\pm$ 2.956 &   2.59 $\pm$  1.27 \cr
H 10$\rightarrow$7 & 8.760 &
 8.760 $\pm$ 0.0012 &  5.75 $\pm$ 1.054 &   3.81 $\pm$  0.93 \cr
H 14$\rightarrow$8 & 8.665  &
 8.666 $\pm$ 0.0018 &  7.48 $\pm$ 1.604 &   3.92 $\pm$  1.12 \cr
H 24$\rightarrow$9 & 8.594 &
 8.592 $\pm$ 0.0026 &                  \nodata  &    \nodata  \cr
H 15$\rightarrow$8 & 8.155 &
 8.158 $\pm$ 0.0009 &  6.49 $\pm$ 0.814 &   6.58 $\pm$  1.09 \cr
H 29$\rightarrow$9 & 8.173& \multicolumn{3}{c}{Blended with 15$\rightarrow$8} \cr
\enddata
\end{deluxetable}

\makeatletter
\def\jnl@aj{AJ}
\ifx\revtex@jnl\jnl@aj\let=tablebreak=\nl\fi
\makeatother

\renewcommand{\arraystretch}{0.75}

\begin{deluxetable}{llllc}
\tablewidth{0pc}
\tablenum{3A}
\tablecaption{Data on observed lines -- $\beta$ Lyr}
\tablehead{
\colhead{Transition\tablenotemark{a}} &
\colhead{$\lambda _{theory} $} &
\colhead{$\lambda _{obs}$} &
\colhead{$\Delta\lambda$}               &
\colhead{$F_{\nu}$}	\\
\colhead{}  	&
\colhead{ (\micron )} &
\colhead{ (\micron )}              &
\colhead{($10^{-3}$\micron )}		&
\colhead{($10^{-19}$ W cm$^{-2}$ \micron $^{-1}$)}
}
\startdata
[Ne II] & 12.81 &
12.813 $\pm$ 0.0017 &  9.91 $\pm$ 1.565 &   8.83 $\pm$  1.84 \cr
H 14$\rightarrow$9 & 12.587 &
12.583 $\pm$ 0.0039 & 10.60 $\pm$ 3.615 &   5.08 $\pm$  2.26 \cr
H 11$\rightarrow$8 & 12.387 &
12.393 $\pm$ 0.0035 &  6.84 $\pm$ 3.384 &   4.09 $\pm$  2.31 \cr
H 7$\rightarrow$6 & 12.372 &
12.368 $\pm$ 0.0020 & 10.16 $\pm$ 1.851 &   38.90 $\pm$  8.70 \cr
H 9$\rightarrow$7 & 11.309 &
11.305 $\pm$ 0.0004 &  9.86 $\pm$ 0.389 &   20.81 $\pm$  1.09 \cr
He I 4P-4S, 3Po-3S & 10.882 &
10.882 $\pm$ 0.0010 &  8.69 $\pm$ 0.953 &   14.24 $\pm$  2.07 \cr
H 16$\rightarrow$9 & 10.804 &
10.801 $\pm$ 0.0071 &  6.85 $\pm$ 6.682 &   2.18 $\pm$  2.65 \cr
H 12$\rightarrow$8 & 10.504 &
10.504 $\pm$ 0.0014 & 10.55 $\pm$ 1.306 &   14.23 $\pm$  2.34 \cr
H 17$\rightarrow$9 & 10.261 &
10.263 $\pm$ 0.0066 &  7.37 $\pm$ 5.992 &   2.45 $\pm$  2.58 \cr
H 18$\rightarrow$9 & 9.847 &
 9.846 $\pm$ 0.0028 &  8.62 $\pm$ 2.576 &   3.12 $\pm$  1.25 \cr
? & \nodata &
 9.597 $\pm$ 0.0012 &  4.34 $\pm$ 1.562 &   2.49 $\pm$  1.12 \cr
H 19$\rightarrow$9 & 9.522 &
 9.516 $\pm$ 0.0027 &  5.74 $\pm$ 2.232 &   2.70 $\pm$  1.44 \cr
H 13$\rightarrow$8 & 9.392 &
 9.389 $\pm$ 0.0012 &  9.77 $\pm$ 1.029 &   10.31 $\pm$  1.43 \cr
H 10$\rightarrow$7 & 8.760 &
 8.758 $\pm$ 0.0007 &  8.14 $\pm$ 0.608 &   23.92 $\pm$  2.36 \cr
H 14$\rightarrow$8 & 8.665  &
 8.663 $\pm$ 0.0030 &  7.64 $\pm$ 2.760 &   7.68 $\pm$  3.53 \cr
H 15$\rightarrow$8 & 8.155 &
 8.153 $\pm$ 0.0020 &  5.11 $\pm$ 1.780 &   6.37 $\pm$  2.75 \cr
\tablenotetext{a}{The hydrogen recombination lines may also be blended with the identical He I recombination transitions.}
\enddata
\end{deluxetable}

\makeatletter
\def\jnl@aj{AJ}
\ifx\revtex@jnl\jnl@aj\let=tablebreak=\nl\fi
\makeatother

\renewcommand{\arraystretch}{0.75}

\begin{deluxetable}{llllc}
\tablewidth{0pc}
\tablenum{3B}
\tablecaption{Data on observed lines -- MWC 349}
\tablehead{
\colhead{Transition} 	&
\colhead{$\lambda _{theory}$} &
\colhead{$\lambda _{obs}$}              &
\colhead{$\Delta\lambda$}               &
\colhead{$F_{\nu}$} \\
\colhead{}  &
\colhead{ (\micron )} &
\colhead{ (\micron )}              &
\colhead{($10^{-3}$\micron )}           &
\colhead{($10^{-17}$ W cm$^{-2}$ \micron $^{-1}$)}
}
\startdata
[Ne II] & 12.81 &
12.813 $\pm$ 0.0002 &  9.55 $\pm$ 0.168 &   95.35 $\pm$  2.21 \cr
H 7$\rightarrow$6 & 12.372 &
12.372 $\pm$ 0.0017 & 13.35 $\pm$ 1.534 &   10.80 $\pm$  1.64 \cr
? & \nodata &
11.918 $\pm$ 0.0045 & 14.39 $\pm$ 4.390 &   5.89 $\pm$  2.32 \cr
H 9$\rightarrow$7 & 11.309 &
11.310 $\pm$ 0.0008 &  7.83 $\pm$ 0.771 &   0.82 $\pm$  0.70 \cr
H 12$\rightarrow$8 & 10.504 &
10.508 $\pm$ 0.0025 &  9.98 $\pm$ 2.329 &   0.06 $\pm$  1.18 \cr
\enddata
\end{deluxetable}

\begin{table}
\tablenum{4}
\begin{center}
\caption{Properties of our Sources}
\begin{tabular}{lccccl}\hline\hline
Object & Shell  & Spectroscopic & X-Ray & Radio & References \\
        & Star &   Binary & Binary & Source &\\\tableline
$\psi$ Per & yes & no & no & yes & 1, 7 \\
$\eta$ Tau & yes & yes & no & yes & 1, 3, 9 \\
$\zeta$ Tau & yes & yes & no & no & 1, 3 \\
48 Per & no & yes & yes & no & 2, 6 \\
$\phi$ Per & yes & yes & no & no & 1 \\
$\gamma$ Cas & yes & no & yes & yes & 1, 4, 9 \\
$o$ Aqr & yes & no & no & no & 1 \\
EW Lac & yes & no & no & yes & 1, 9 \\
$\kappa$ Dra & no & yes & yes & no & 5, 6 \\
$\beta$ Lyr & no & yes & no & yes & 5, 10 \\
MWC 349 & no & no & no & yes & 8 \\
\end{tabular}
\tablenotetext{}{(1) Slettebak {\it et al.} 1992, (2) Poeckert 1981, (3) Jared {\it et al.} 1989, (4) Haberl 1995, (5) Pols {\it et al.} 1991, (6) Peters 1982, (7) Marlborough {\it et al.} 1997, (8) Strelnitski {\it et al.} 1996, (9) Taylor {\it et al.} 1990, (10)Pe\u{s}ek {\it et al.} 1996.}
\end{center}
\end{table}

\makeatletter
\def\jnl@aj{AJ}
\ifx\revtex@jnl\jnl@aj\let=tablebreak=\nl\fi
\makeatother

\renewcommand{\arraystretch}{0.75}

\begin{deluxetable}{lllllll}
\tablewidth{0pc}
\tablenum{5}
\tablecaption{Stellar Parameters for Sources}
\tablehead{
\colhead{Object}                    &
\colhead{$T_{eff}$}           &
\colhead{$R_{\star}$}      &
\colhead{$D$}			&
\colhead{$\log L$}			&
\colhead{1.25\micron}	&
\colhead{2.2\micron}\\
\colhead{}				&
\colhead{(K)}				&
\colhead{(R$_{\sun}$)}			&
\colhead{(pc)}				&
\colhead{($\log L_{\sun}$)}		&
\colhead{(mag)}				&
\colhead{(mag)}
}
\startdata
$\psi$~Per    & 16000 & 7.0 & 215 & 3.20 & 4.30 & 3.94\\
$\eta$~Tau    & 14000 & 6.5 & 113 & 2.86 & 2.95 & 2.96\\
$\zeta$~Tau   & 16500 & 6.5 & 128 & 3.05 & 3.20 & 2.94\\
48 Per        & 16000 & 6.3 & 170 & 3.60 & 3.99\tablenotemark{a} & 3.79\tablenotemark{b}\\
$\phi$~Per    & 20000 & 7.5 & 220 & 3.20 & 3.83 & 3.25\\
$\gamma$~Cas  & 31000 & 10.0 & 188 & 4.93 & 2.47 & 2.20\\
$o$ Aqr       & 11000 & 4.0 & 117 & 2.45 & 4.84 & 4.80 \\
EW Lac        & 16000 & 6.5 & 337 & 3.54 & 5.34 & 5.00\\
$\kappa$~Dra  & 14500 & 6.4 & 153 & 2.70 & 3.86 & 3.75
\tablenotetext{a}{At 1.1\micron .}
\tablenotetext{b}{At 2.3\micron .}
\enddata
\end{deluxetable}


\makeatletter
\def\jnl@aj{AJ}
\ifx\revtex@jnl\jnl@aj\let=tablebreak=\nl\fi
\makeatother

\renewcommand{\arraystretch}{0.75}

\begin{deluxetable}{lcccc}
\tablewidth{0pc}
\tablenum{6}
\tablecaption{Parameter Values for Free-Free Fits}
\tablehead{
\colhead{Object}                    &
\colhead{$T_{sh}$}           &
\colhead{$n_e$}	 	&	
\colhead{$R_{sh}$}      &
\colhead{$\chi ^2$}\\
\colhead{}				&
\colhead{(10$^4$ K)}				&
\colhead{(10$^{11}$ cm$^{-3}$)}		&
\colhead{(R$_{\star}$)}			&
\colhead{}		
}
\startdata
$\psi$~Per    & 1.60 (+.07/-.15) & 2.60 (+.16/-.07) & 4.61 (+.21/-.20) & 1.56  \\
$\eta$~Tau    & 0.80 (+.02/-.06) & 1.52 (+.08/-.03) & 3.07 ($\pm$.15)  & 1.04\\
$\zeta$~Tau   & 1.39 (+.04/-.10) & 2.30 (+.11/-.04) & 4.95 ($\pm$.32)  & 4.27\\
48 Per        & 1.29 (+.05/-.08) & 2.21 (+.09/-.05) & 4.04 ($\pm$.13)  & 1.05\\
$\phi$~Per    & 1.65 (+.06/-.08) & 3.01 (+.19/-.10) & 5.02 (+.10/-.11) & 0.74\\
$\gamma$~Cas  & 2.68 (+.15/-.14) & 4.57 (+.33/-.21) & 6.00 ($\pm$.13)  & 1.12\\
$o$ Aqr       & 1.01 (+.02/-.11) & 1.60 (+.05/-.03) & 2.99 (+.11/-.13) & 0.54\\
EW Lac        & 1.32 (+.02/-.09) & 2.20 (+.08/-.03) & 3.80 (+.20/-.22) & 1.49\\
$\kappa$~Dra  & 1.20 (+.04/-.11) & 1.83 (+.14/-.04) & 3.12 (+.14/-.16) & 2.13\\
\enddata
\end{deluxetable}

\makeatletter
\def\jnl@aj{AJ}
\ifx\revtex@jnl\jnl@aj\let=tablebreak=\nl\fi
\makeatother

\renewcommand{\arraystretch}{0.75}

\begin{deluxetable}{llllllll}
\tablewidth{0pc}
\tablenum{7}
\tablecaption{Comparision of Free-Free Fits}
\tablehead{
\colhead{Object}                    &
\colhead{Author\tablenotemark{a}}		&
\colhead{$T_{\star}$}           &
\colhead{$R_{\star}$}      &
\colhead{$T_{sh}$}	 		&
\colhead{$R_{sh}$}				&
\colhead{$n_e$}				&
\colhead{M$_H$}				\\
\colhead{}				&
\colhead{}				&
\colhead{(10$^4$ K)}			&
\colhead{(R$_{\sun}$)}			&
\colhead{(10$^4$ K)}			&
\colhead{(R$_{\sun}$)}			&
\colhead{(10$^{11}$ cm$^{-3}$)}		&
\colhead{(10$^{-10}$ M$_{\sun}$)}
}
\startdata
$\psi$~Per    & RHS & 1.6 & 7.0 & 1.6 & 32.3 & 2.60 & 9.3 \\
	      & WCL & 1.5 & 6.0 & 1.2 & 37.2 & \nodata\\
$\eta$~Tau    & RHS & 1.4 & 6.5 & 0.8 & 20.0 & 1.52 & 0.93\\
$\zeta$~Tau   & RHS & 1.65 & 6.5 & 1.39 & 32.2 & 2.30 & 8.3\\
	      & GHJ & 1.8 & 5.75 & 1.4 & 20.1 & 4.4\\
	      & WCL & 1.4 & 6.0 & 1.1 & 28.8 & \nodata\\
48 Per        & RHS & 1.6 & 6.3 & 1.29 & 25.5 & 2.21 & 3.7\\
	      & WCL & 1.8 & 6.5 & 1.45 & 36.4 & \nodata \\
$\phi$~Per    & RHS & 2.0 & 7.5 & 1.65 & 37.7 & 3.01 & 17.6\\
	      & GHJ & 1.6 & 5.0 & 1.2 & 23.0 & 3.8 \\
	      & WCL & 2.0 & 7.0 & 1.6 & 91.0 & \nodata\\
$\gamma$~Cas  & RHS & 3.1 & 9.0 & 2.68 & 54.0 & 4.57 & 81.4\\
	      & GHJ & 2.6 & 7.2 & 2.0 & 29.5 & 4.4 \\
	      & WCL & 3.1 & 10.0 & 2.5 & 80.0 & \nodata \\ 
$o$ Aqr       & RHS & 1.1 & 4.0 & 1.01 & 12.0 & 1.9 & 0.25\\
	      & WCL & 1.3 & 3.4 & 1.05 & 9.5 &\nodata \\
EW Lac        & RHS & 1.6 & 6.5 & 1.32 & 24.7 & 2.20 & 3.2\\
	      & WCL & 1.4 & 5.7 & 1.1 & 27.4 & \nodata\\
$\kappa$~Dra  & RHS & 1.45 & 6.5 & 1.20 & 20.3 & 1.83 & 1.2\\
	      & WCL & 1.3 & 4.5 & 1.05 & 22.1 & \nodata \\
\tablenotetext{a}{WCL = Waters {\it et al.} 1987, GHJ = Gehrz {\it et al} 1974, and RHS = this work.}
\enddata
\end{deluxetable}

\begin{figure}
\plottwo{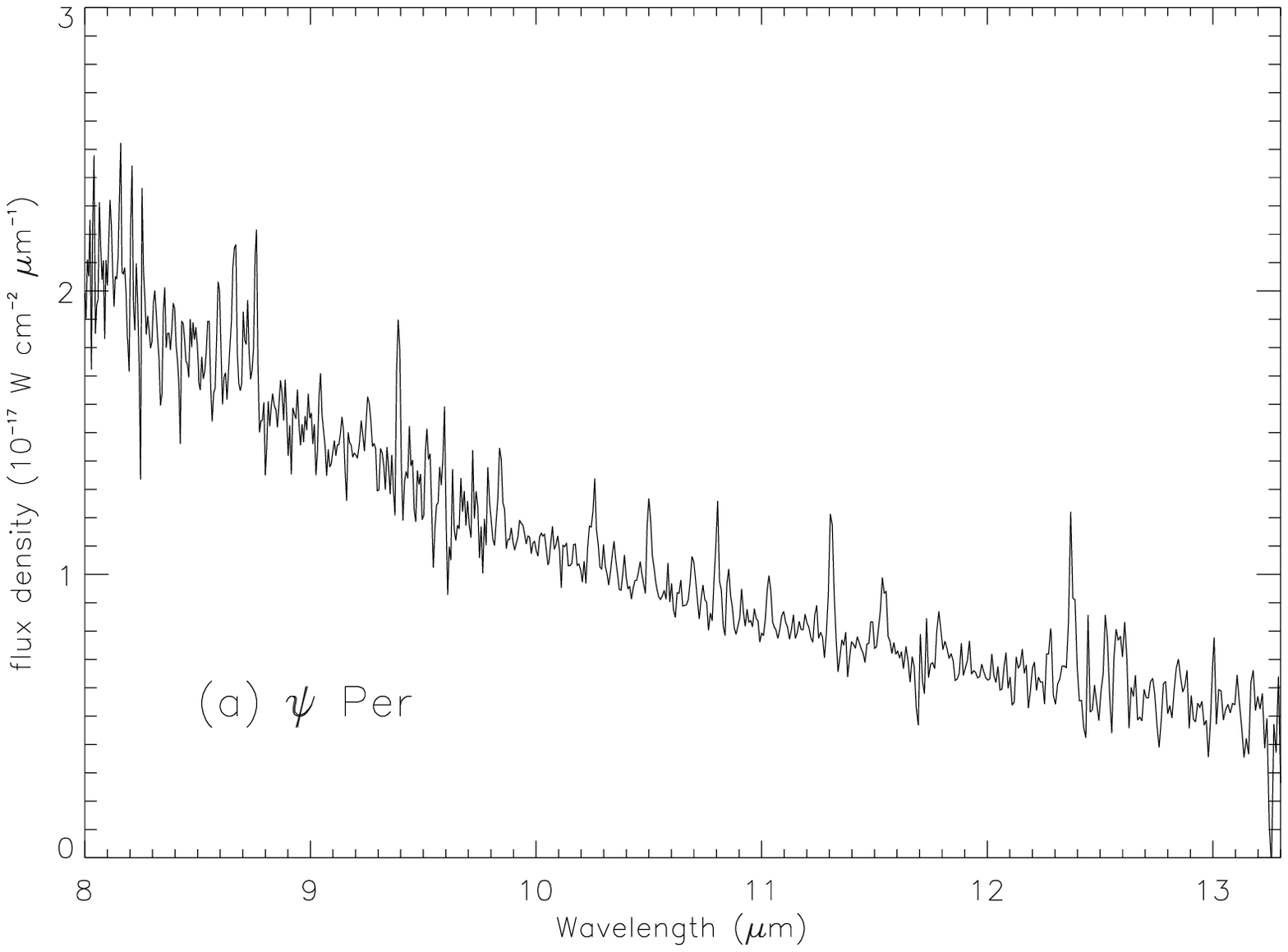}{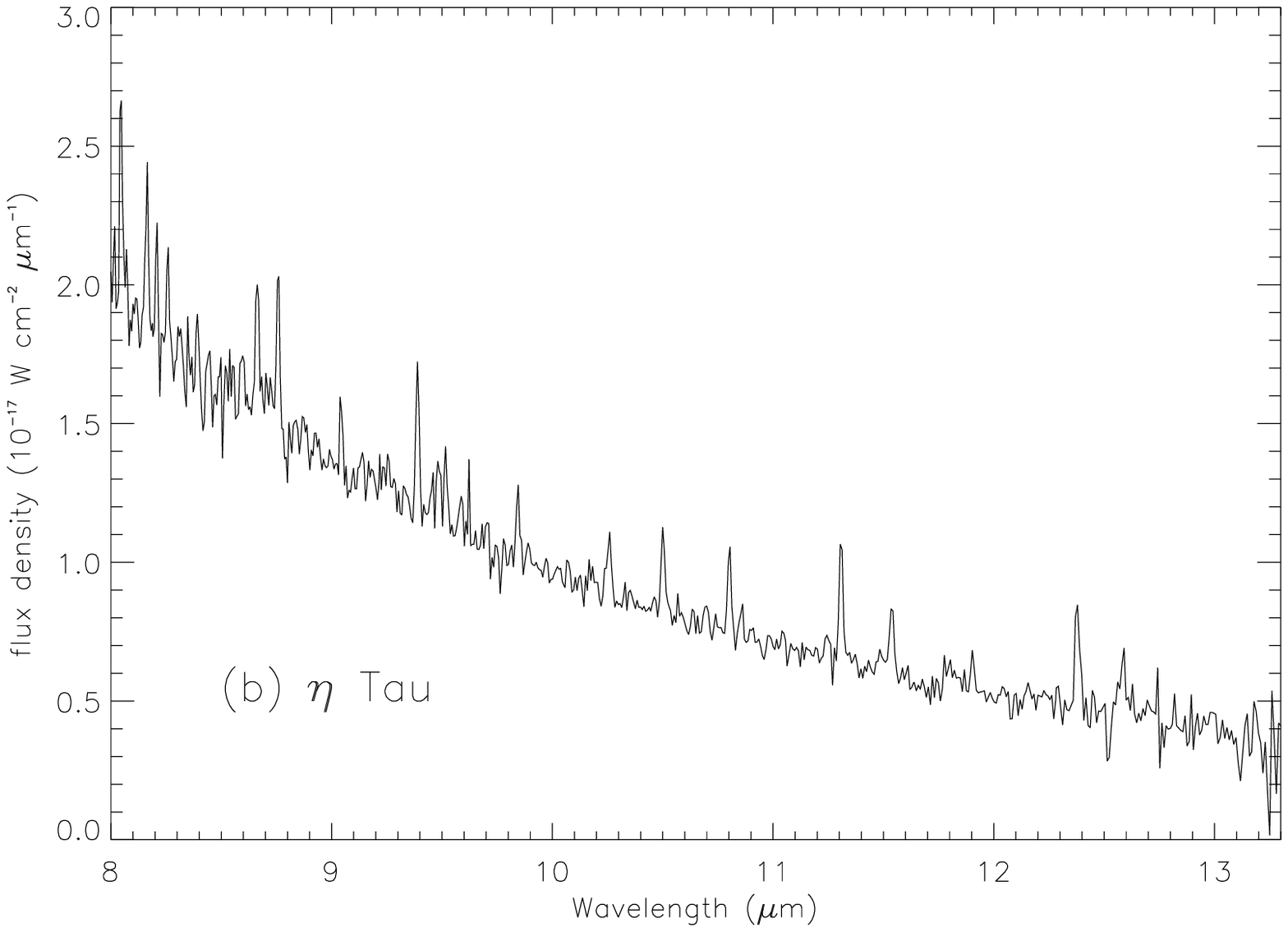}
\end{figure}
\begin{figure}
\plottwo{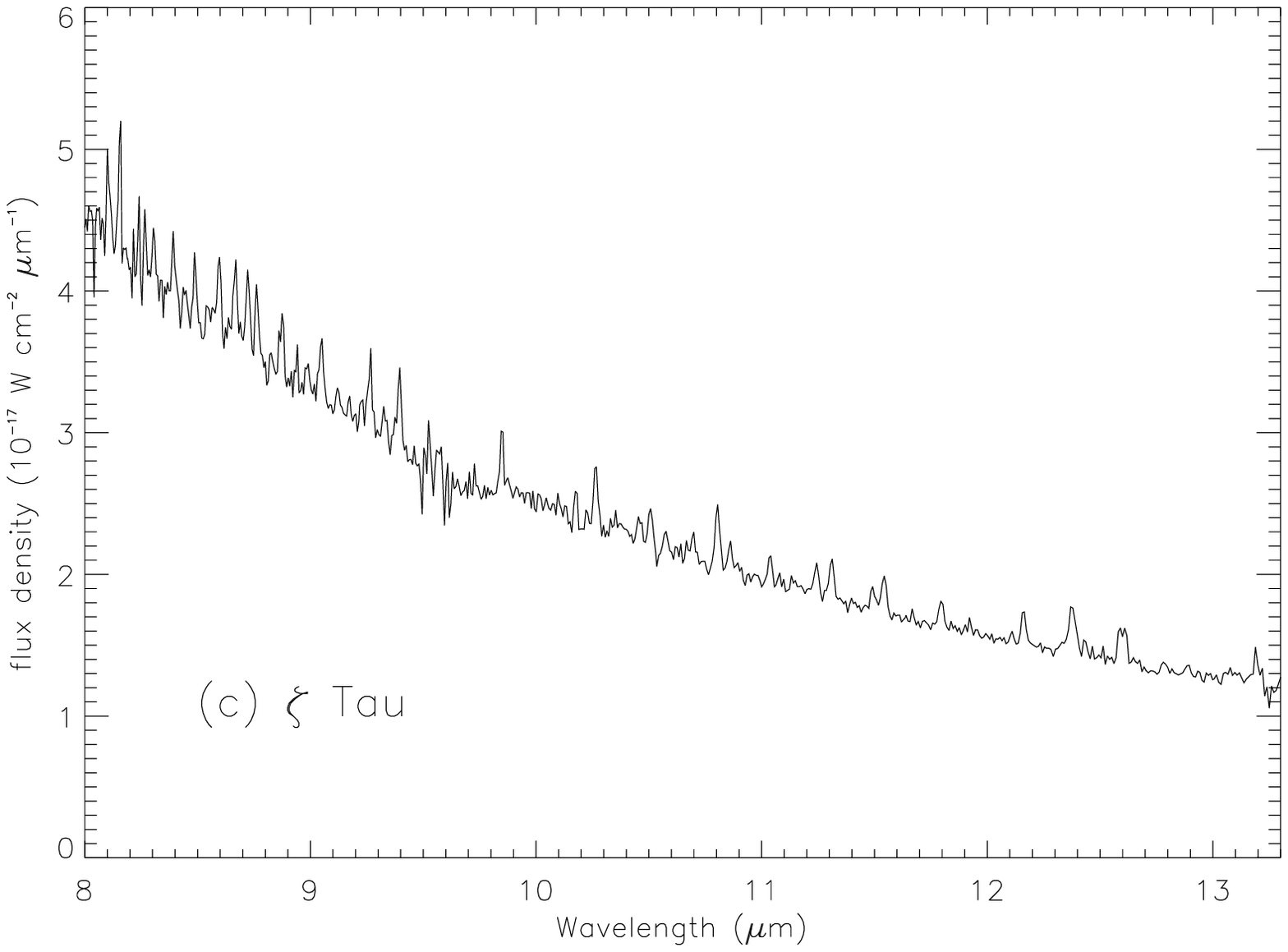}{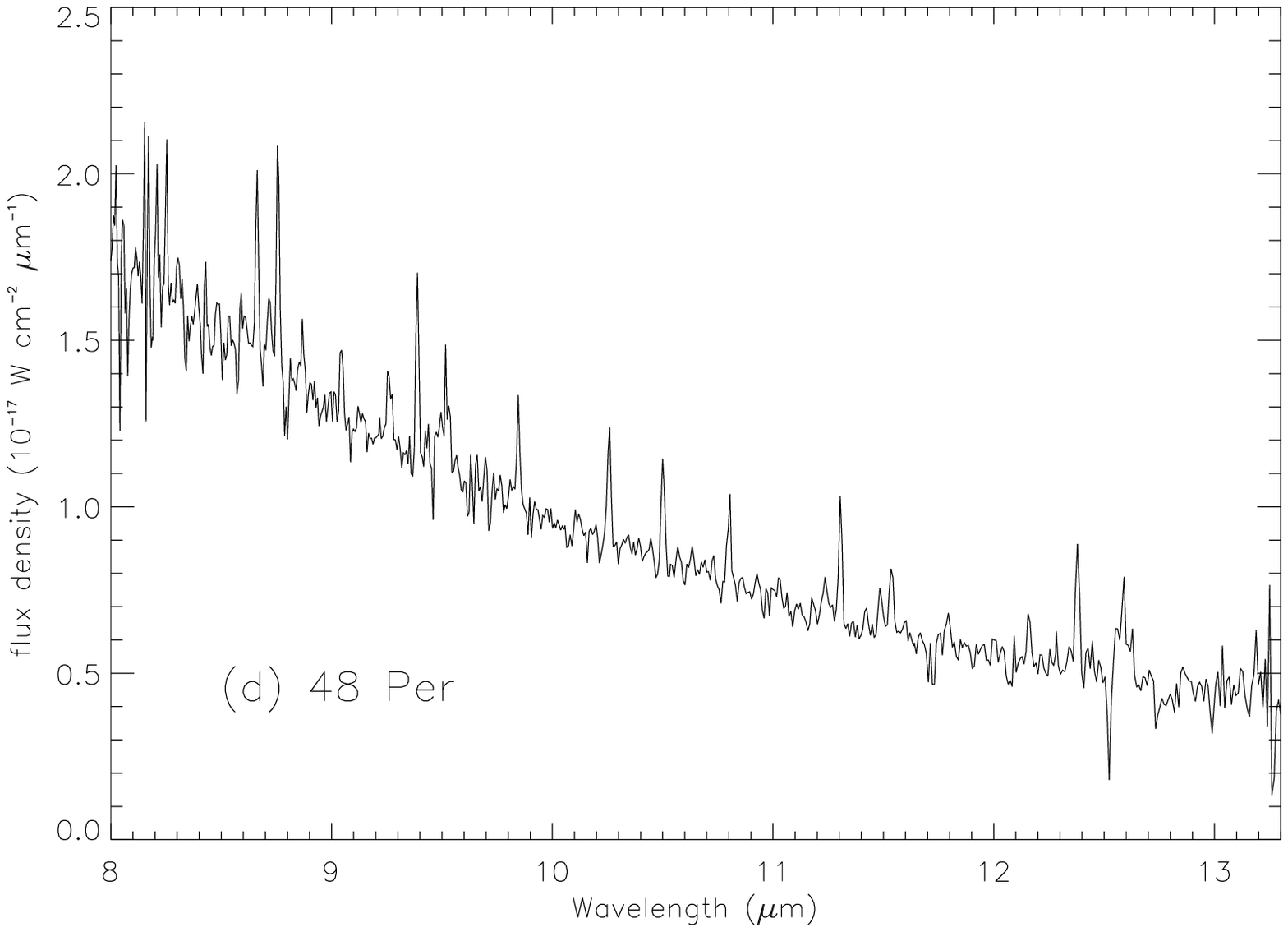}
\end{figure}
\begin{figure}
\plottwo{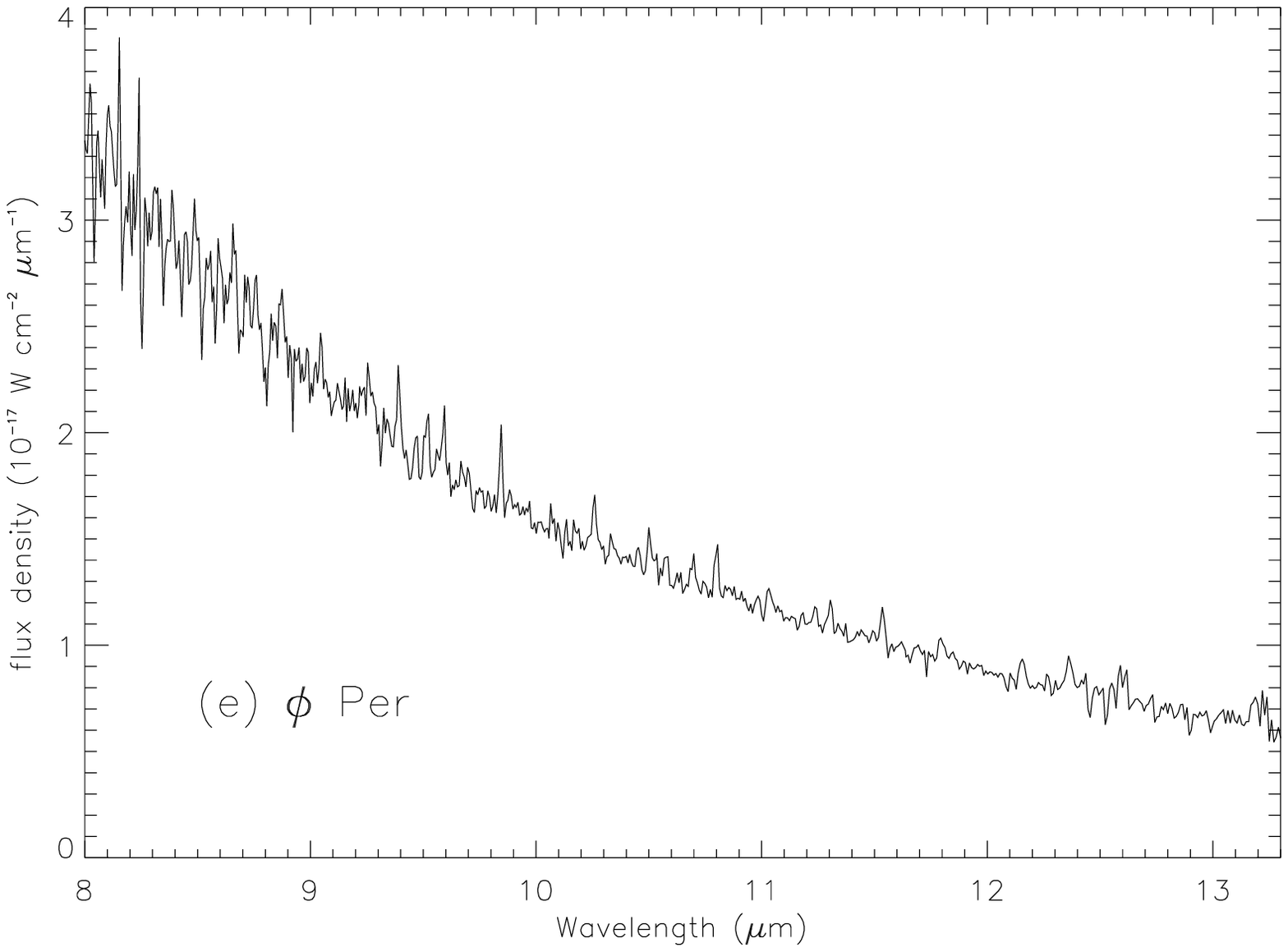}{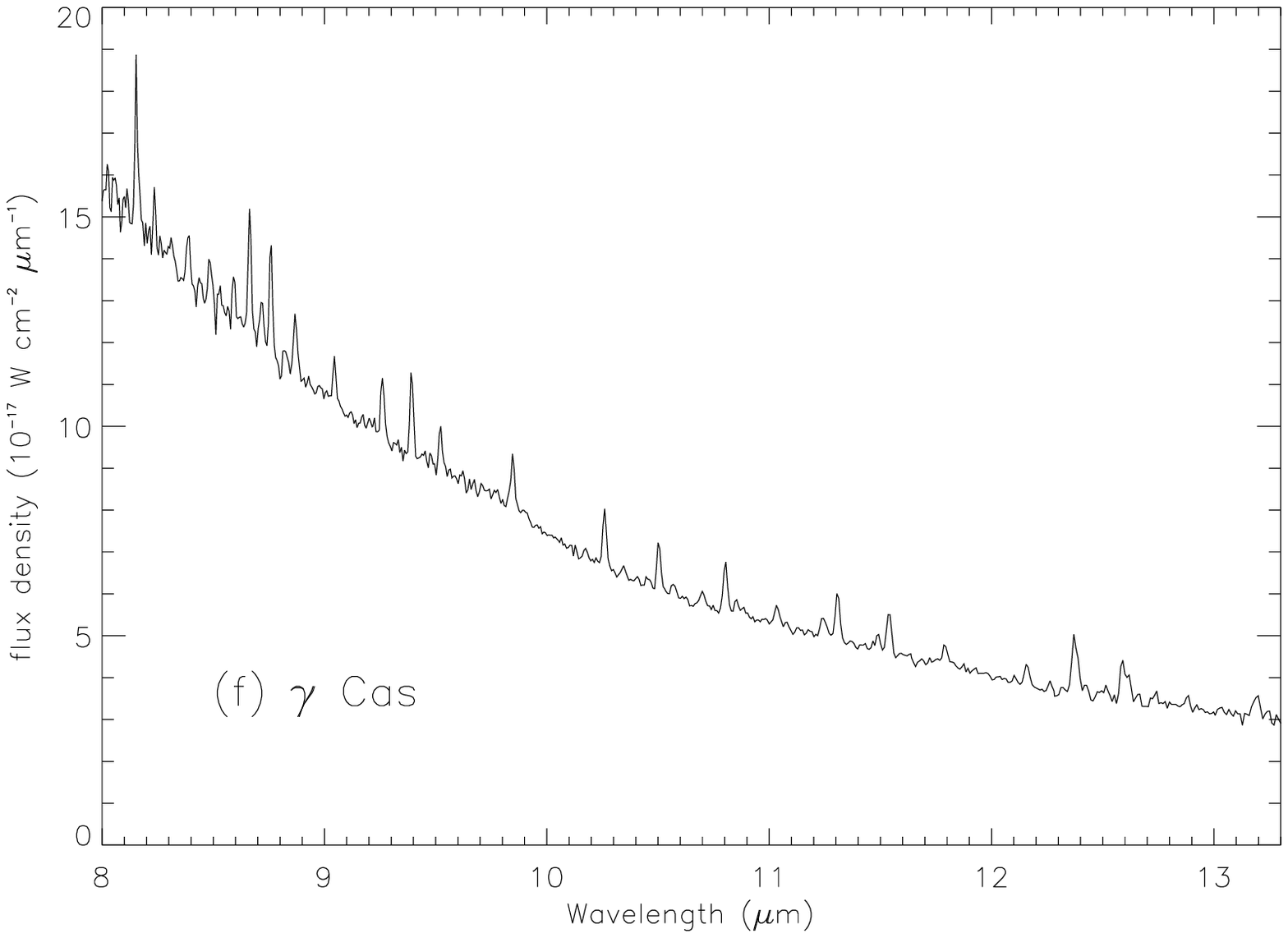}
\caption{The reduced spectra of our sources, (a) $\Psi$~Per, (b) $\eta$~Tau, (c) $\zeta$~Tau, (d) 48~Per, (e) $\phi$~Per, (f) $\gamma$~Cas.}
\end{figure}

\begin{figure}
\plottwo{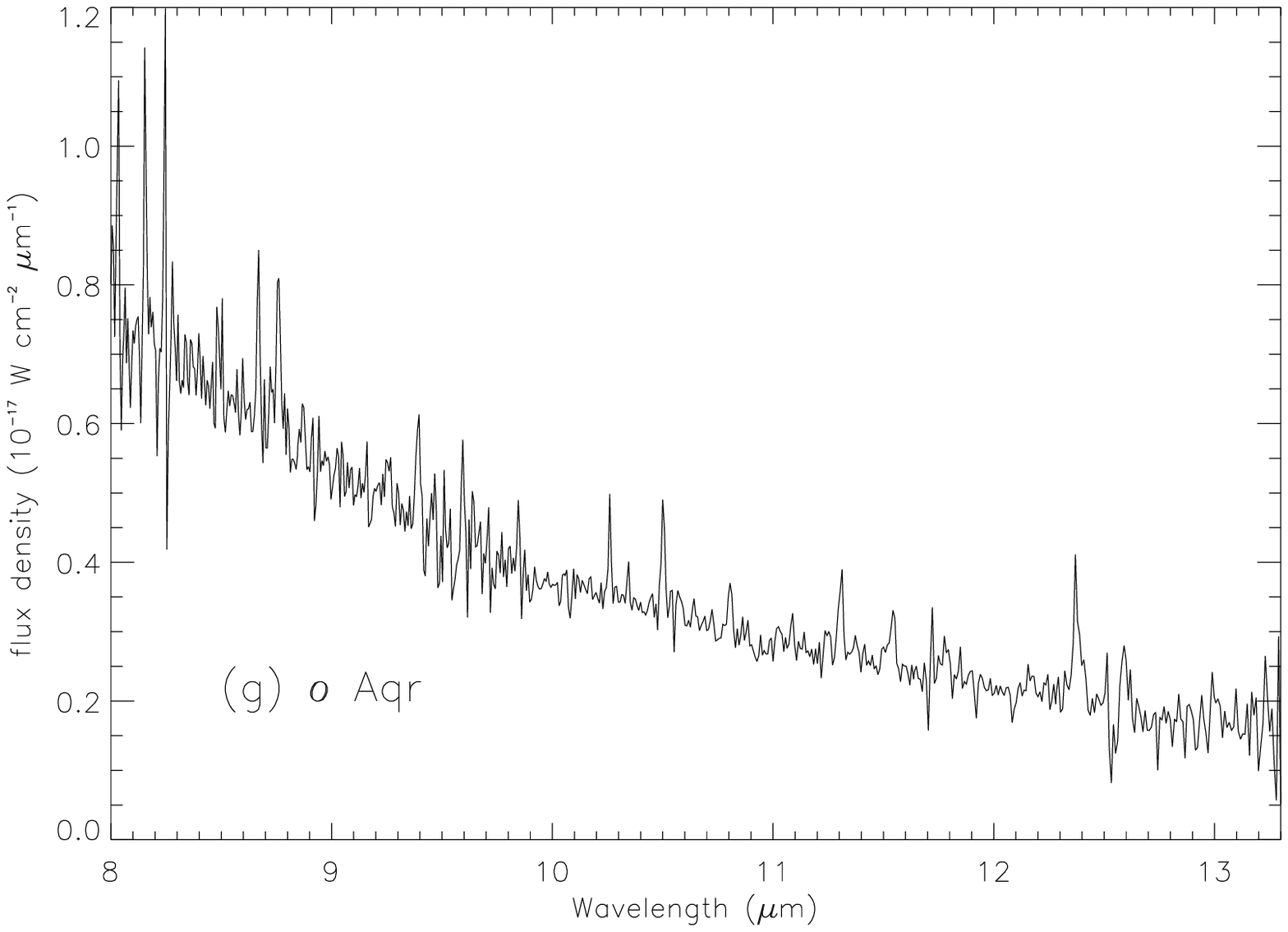}{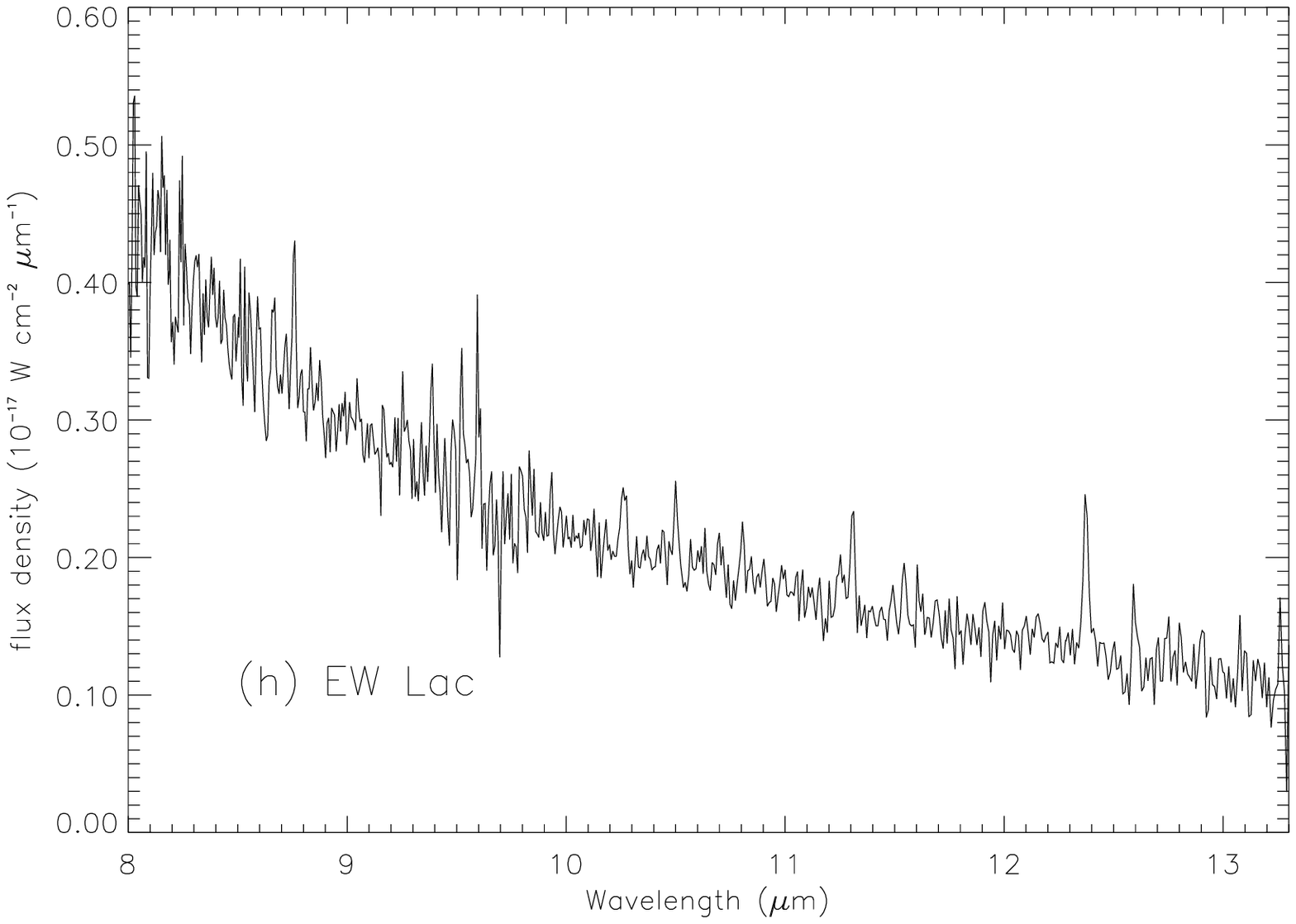}
\end{figure}
\begin{figure}
\plottwo{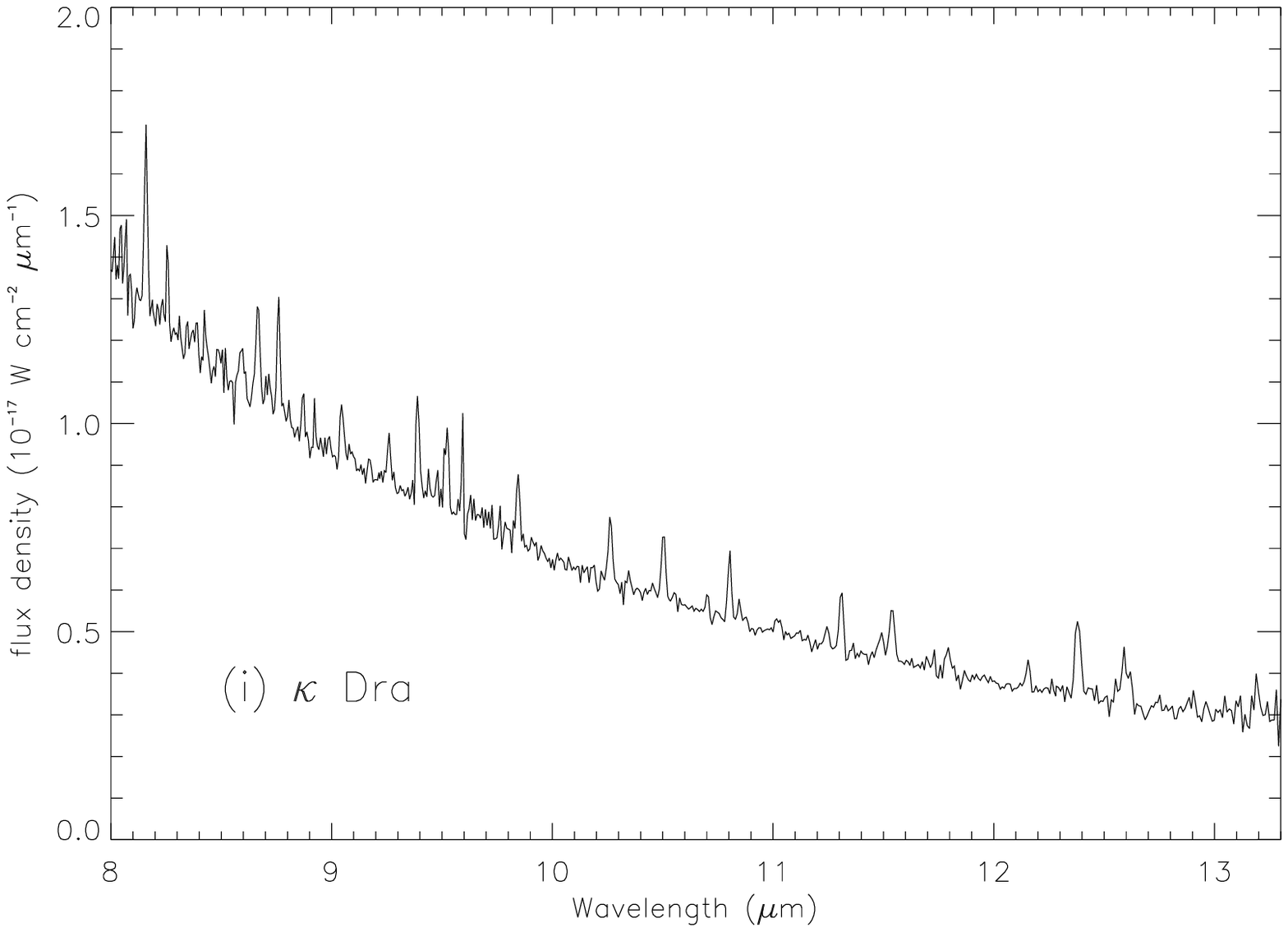}{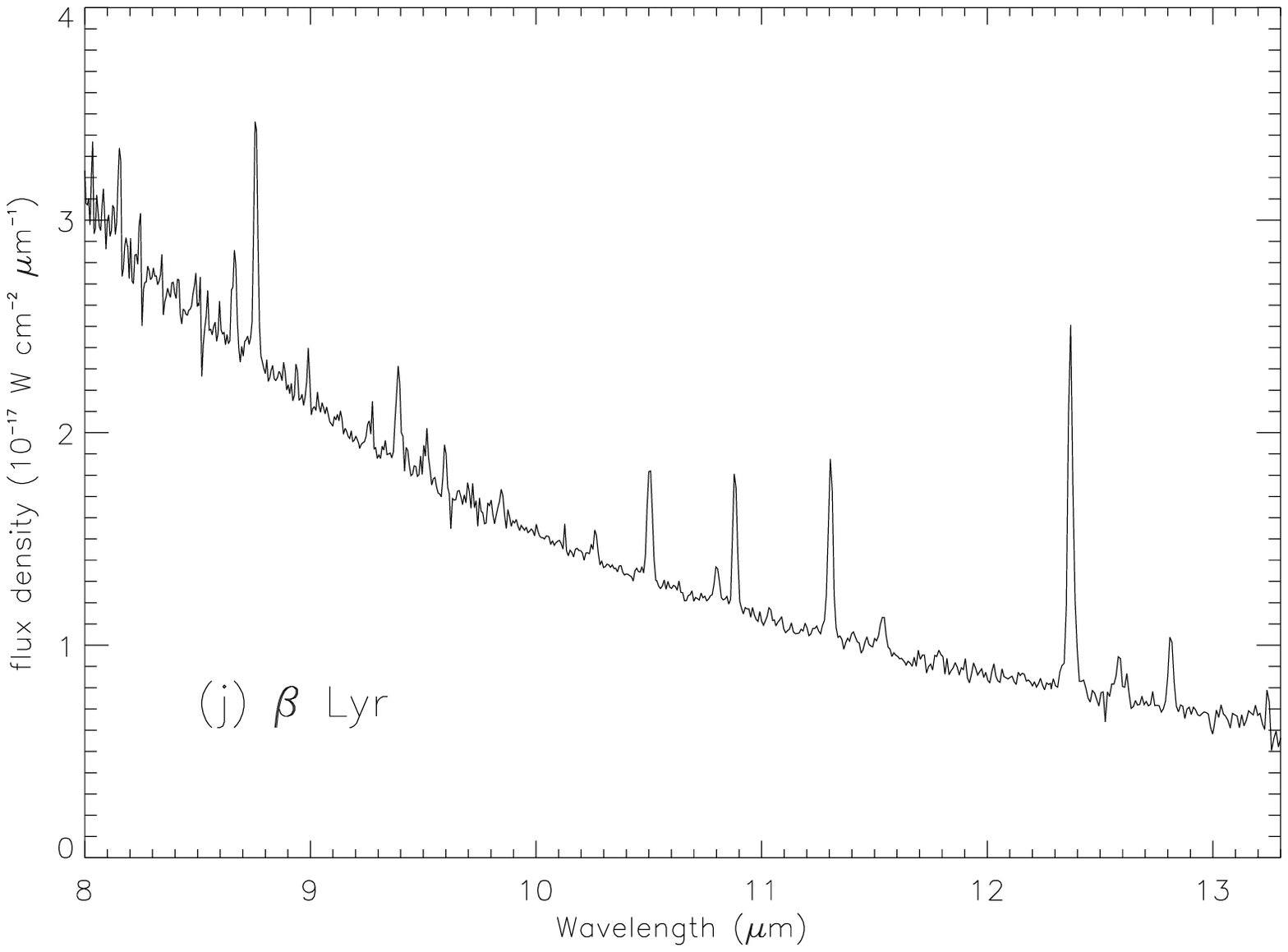}
\end{figure}
\begin{figure}
\figurenum{1, Continued}
\epsscale{0.5}
\plotone{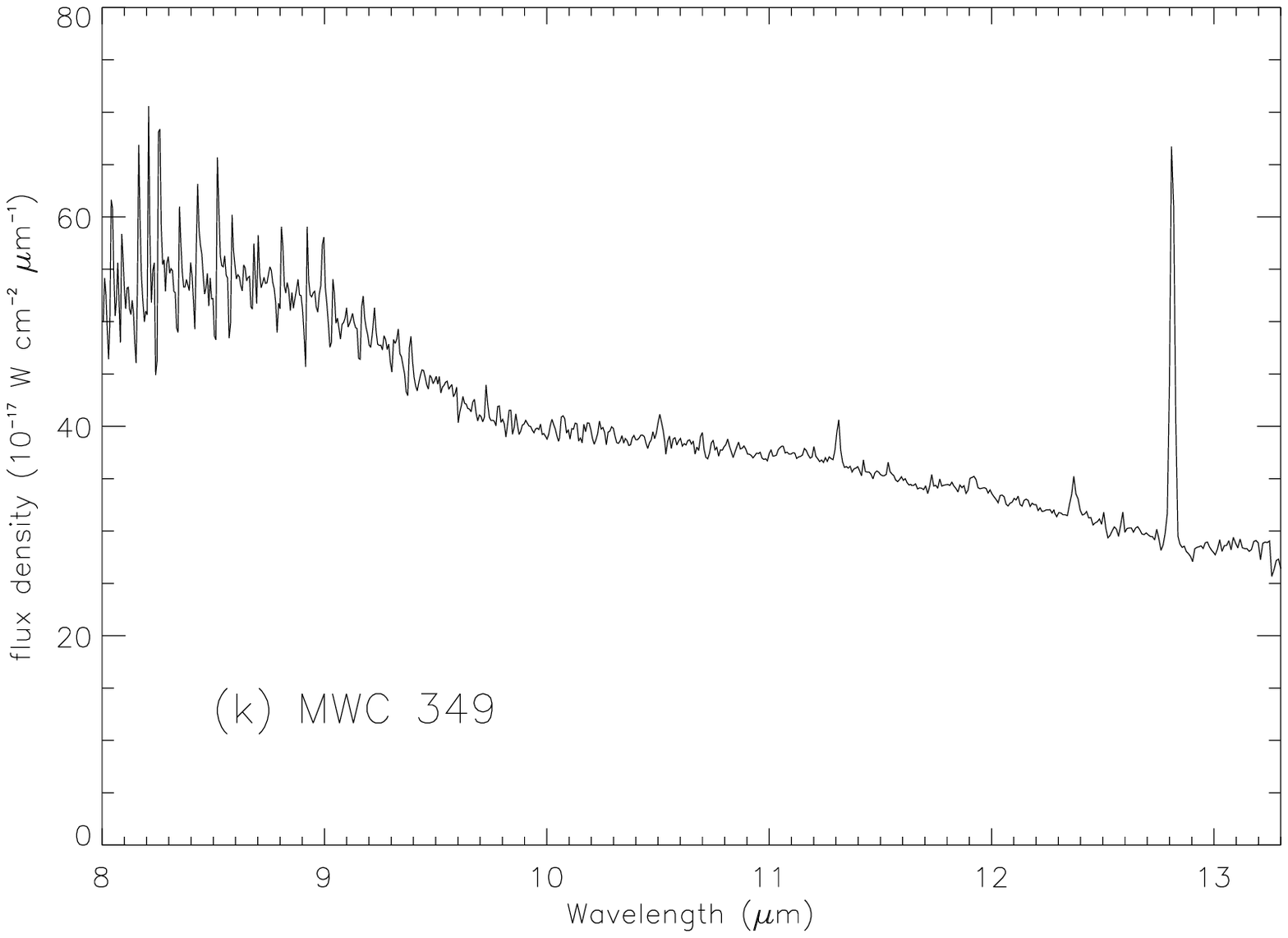}
\caption{The reduced spectra of our sources, continued.  (g) $o$~Aqr, (h) EW~Lac, (i) $\kappa$~Dra, (j) $\beta$~Lyr, and (k) MWC~349.}
\end{figure}

\begin{figure}
\epsscale{0.7}
\plotone{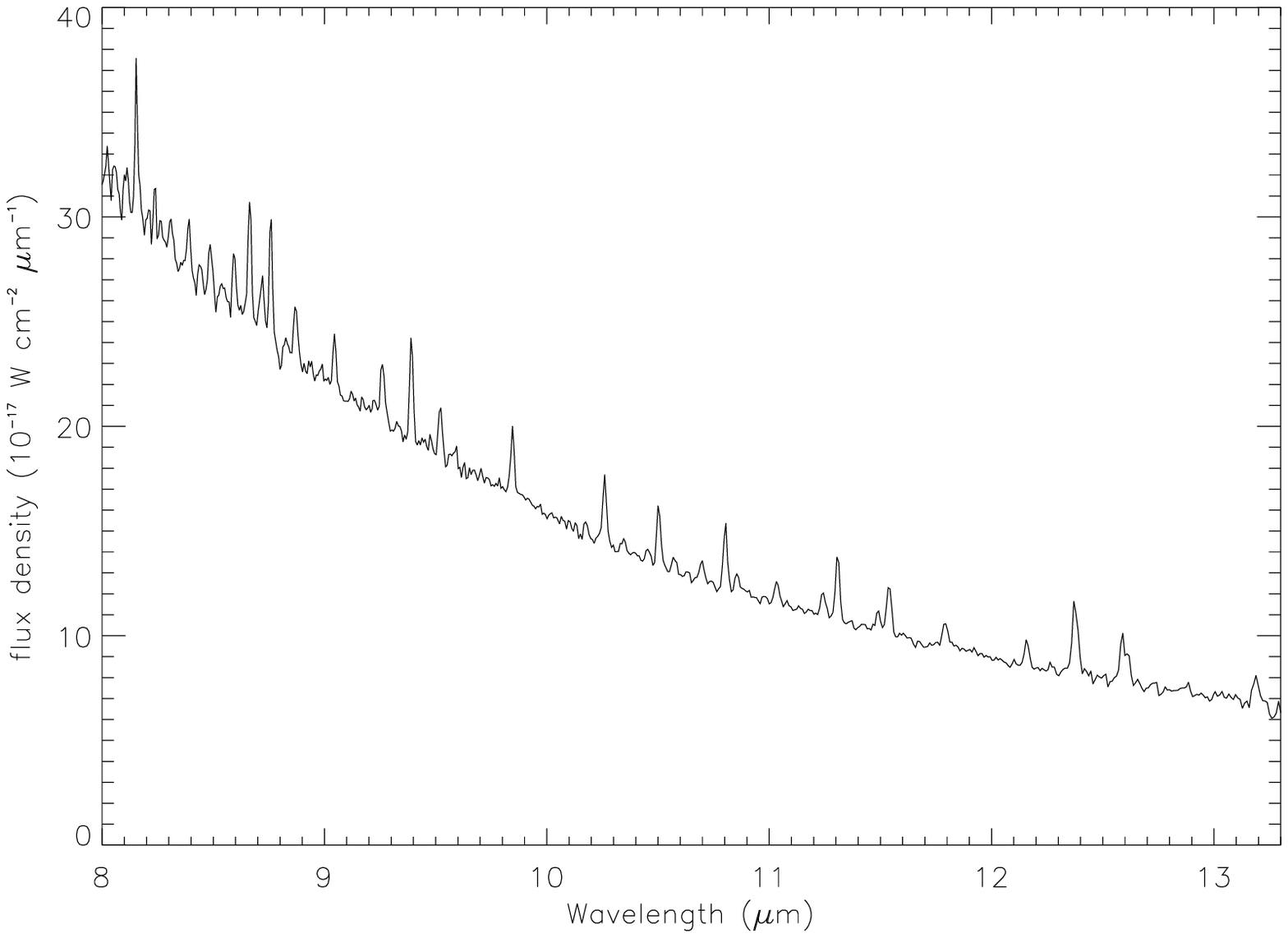}
\caption{The spectrum produced by coadding all nine of the hydrogen spectra (HS) Be star spectra.  Several lines which are visible in the individual spectra become clearer in this summed spectra.}
\end{figure}

\begin{figure}
\epsscale{0.7}
\plotone{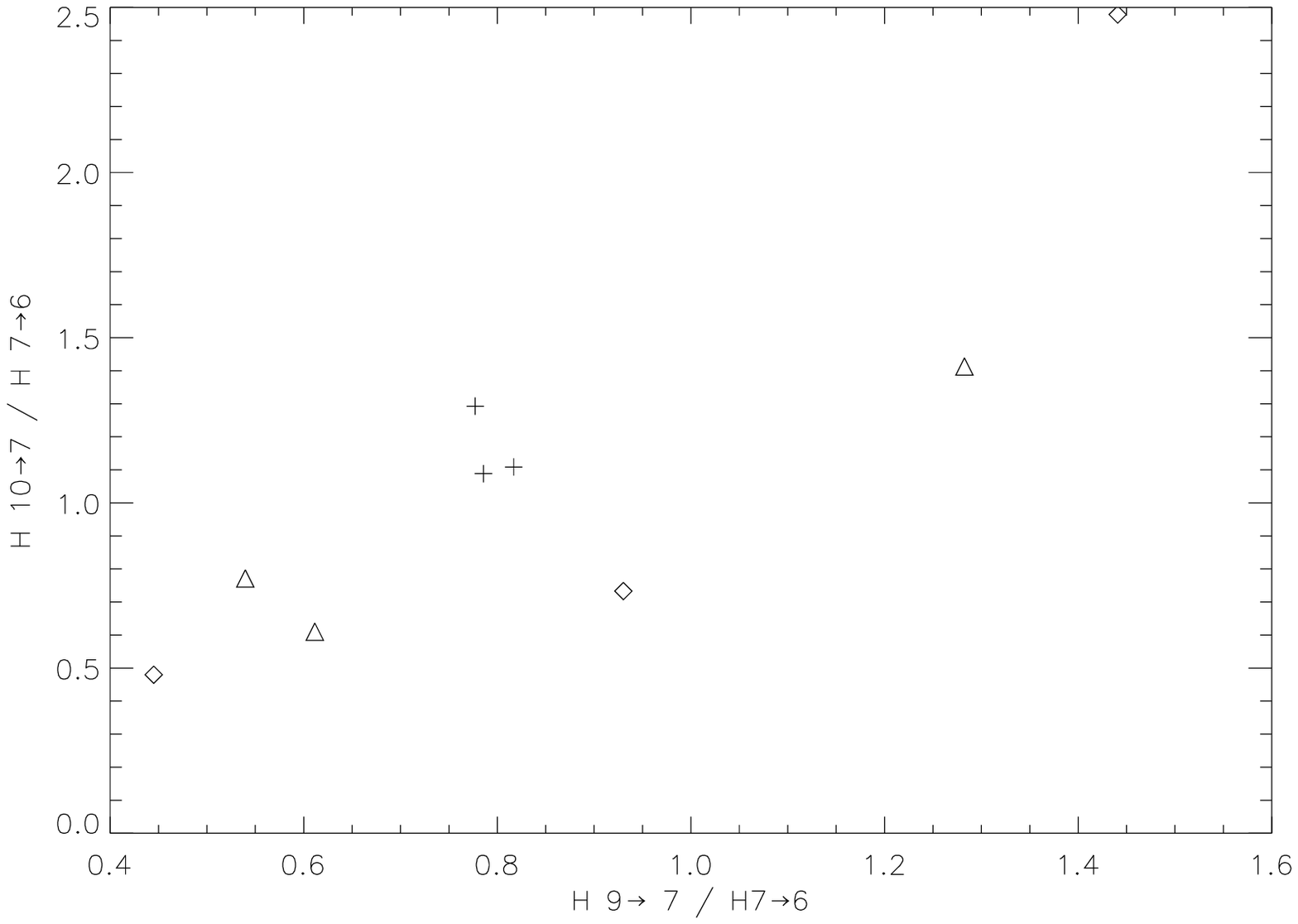}
\caption{A comparison plot of the H 9$\rightarrow$7/H 7$\rightarrow$6 ratio versus the H 10$\rightarrow$7/H 7$\rightarrow$6 ratio.  No strong correlation can be seen in this small sample set.}
\end{figure}

\begin{figure}
\figurenum{4}
\epsscale{0.7}
\plotone{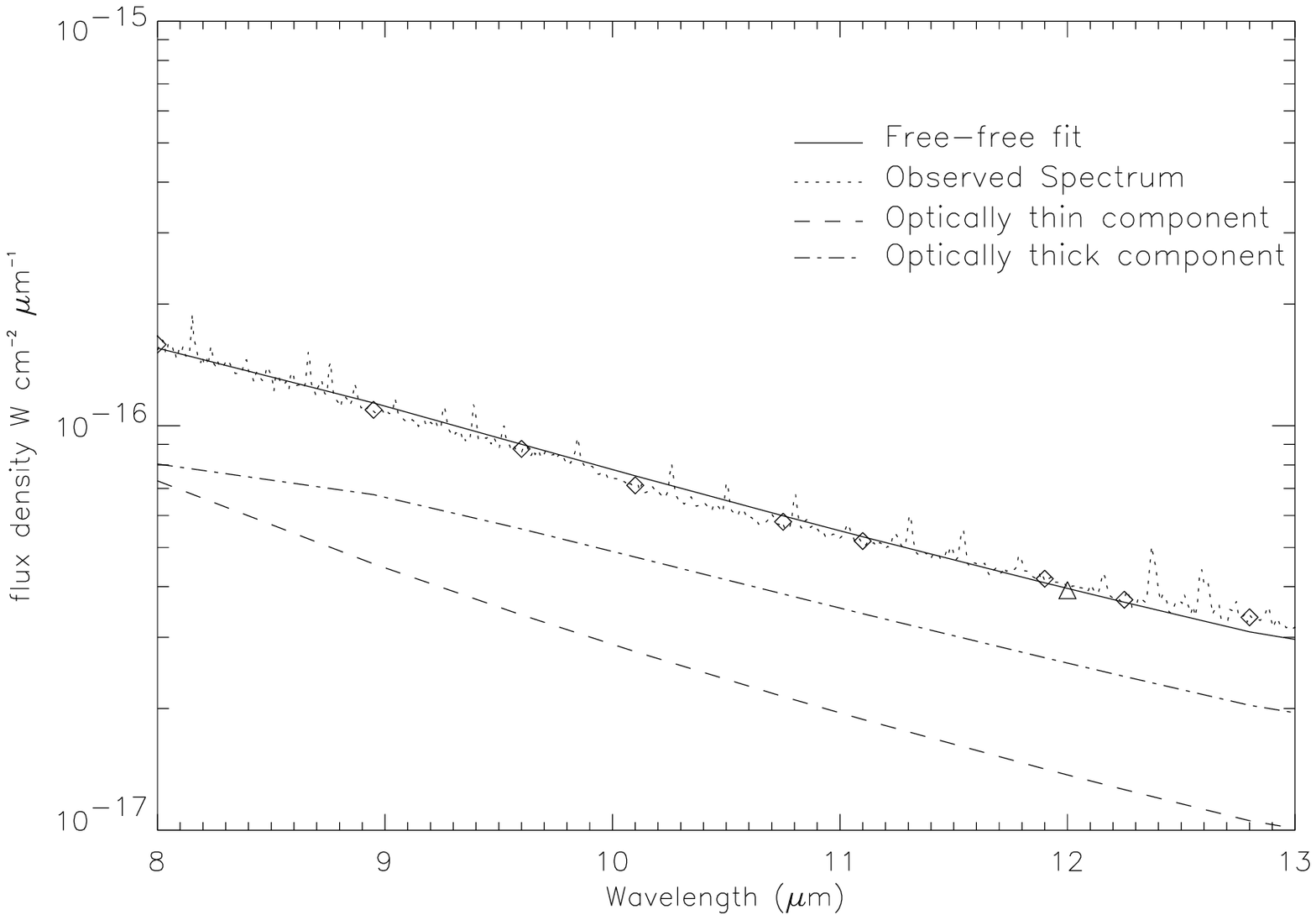}
\caption{The spectrum of $\gamma$~Cas (dotted line), with the free-free emission fit overlaid (solid line).  The dashed line and dash-dot line are the optically thin and optically thick components of the emission, respectively.  The open diamonds are the points used to make the fit, and the open triangle is the IRAS 12\micron \ flux.}
\end{figure}

\end{document}